\documentclass[prb, twocolumn, showpacs,10pt,superscriptaddress]{revtex4-1}

\usepackage{amssymb,amsfonts,amsmath} 
\usepackage{graphicx}
\usepackage{hyperref}
\usepackage{color}
\usepackage{graphics}
\usepackage[dvipsnames]{xcolor}
\usepackage{enumitem}
\usepackage{ulem}

\bibpunct{[}{]}{,}{n}{}{}

\definecolor{lightblue}{RGB}{89, 171, 227}

\definecolor{change}{RGB}{255, 0, 0}

\begin{document} 

\title{
Quantitative analysis of interaction effects in generalized Aubry-Andr\'e-Harper models
}

\author{Y.-T.~Lin}
\affiliation{Institut f{\"u}r Theorie der Statistischen Physik, RWTH Aachen University and
  JARA---Fundamentals of Future Information Technology, 52056 Aachen, Germany}
\author{C.~S.~Weber}
\affiliation{Institut f{\"u}r Theorie der Statistischen Physik, RWTH Aachen University and
  JARA---Fundamentals of Future Information Technology, 52056 Aachen, Germany}
\author{D.~M.~Kennes}
\affiliation{Institut f{\"u}r Theorie der Statistischen Physik, RWTH Aachen University and
  JARA---Fundamentals of Future Information Technology, 52056 Aachen, Germany}
 \affiliation{Max Planck Institute for the Structure and Dynamics of Matter, Center for Free Electron Laser Science, 22761 Hamburg, Germany}
\author{M.~Pletyukhov}
\affiliation{Institut f{\"u}r Theorie der Statistischen Physik, RWTH Aachen University and
  JARA---Fundamentals of Future Information Technology, 52056 Aachen, Germany}
\author{H.~Schoeller}
\affiliation{Institut f{\"u}r Theorie der Statistischen Physik, RWTH Aachen University and
  JARA---Fundamentals of Future Information Technology, 52056 Aachen, Germany}
\author{V.~Meden}
\affiliation{Institut f{\"u}r Theorie der Statistischen Physik, RWTH Aachen University and
  JARA---Fundamentals of Future Information Technology, 52056 Aachen, Germany}

\begin{abstract}
We present a quantitative analysis of two-particle interaction effects in generalized, one-dimensional Aubry-Andr\'e-Harper models with the Fermi energy placed in one of the band gaps. 
We investigate systems with periodic as well as open boundary conditions; for the latter focusing on the number of edge states and the boundary charge. Both these observables are important for the classification of noninteracting topological systems.
In our first class of models the unit cell structure stems from periodically modulated single-particle parameters. In the second it results from the spatial modulation of the two-particle interaction. For both types of models we find that the single-particle band gaps are renormalized by the interaction in accordance with expectations employing general field theoretical arguments. While interaction induced effective edge states can be found in the local single-particle spectral function close to a boundary, the characteristics of the boundary charge are not modified by the interaction.   
This indicates that our results for the Rice-Mele and Su-Schriefer-Heeger model [Phys.~Rev.~B {\bf 102}, 085122 (2020)] are generic and can be found in lattice models with more complex unit cells as well. 
\end{abstract}

\date{\today} 
\maketitle

\section{Introduction}
\label{sec:intro}

In the context of topological properties of fermionic systems, one-dimensional (1D) lattice models with a rich structure of the unit cell have attracted a tremendous interest in condensed matter physics \cite{kane_mele_prl_05,bernevig_etal_science_06,fu_kane_mele_prl_07,koenig_etal_science_07,hsieh_etal_nature_08,hasan_kane_RMP_10,qi_zhang_RMP_11,bernevig_book_13,tkachov_book_15,asboth_book_16}. Besides edge states, i.e., bound states located at the boundary of the system, also the investigation of the boundary charge accumulated at one end of the system together with its fluctuations recently experienced a significant revival \cite{park_etal_prb_16,thakurathi_etal_prb_18,pletyukhov_etal_short,pletyukhov_kennes_klinovaja_loss_schoeller_prb_20,rational_boundary_charge_in_1d_prr_20,weber_etal_prl_21}.
Corresponding models are characterized by several bands which are separated by single-particle gaps. The topological properties as well as the presence or absence of in-gap edge states localized close to an open boundary can be tuned by changing the single-particle parameters, i.e., the hopping matrix elements and the onsite energies. Comparably simple models from this class are the Su-Schriefer-Heeger (SSH) model \cite{Su79,Heeger01} and the Rice-Mele (RM) model \cite{Heeger01,Rice82} with a unit  cell size of $Z=2$ lattice sites. Already shortly after these models were set up in the early eighties effects of a complementary two-particle interaction were investigated. This was mainly done using effective low-energy field theories which are supposed to be applicable to more complex microscopic lattice models as well \cite{Kivelson85,Horovitz85,Wu86,Jackiw83,Jackiw76}.

We recently studied the spinless RM model and, as a special limit of this, the SSH model with a homogeneous nearest-neighbor two-particle interaction directly, i.e., without the approximate mapping to a low-energy continuum field theory \cite{RM_frg_prb_20}. Employing an approximate functional renormalization group (RG) \cite{Metzner12,Kopietz10} approach and numerically exact density matrix renormalization group (DMRG) \cite{Schollwoeck11,J02,BSW09,KKM14} we
\begin{enumerate}
    \item confirmed the low-energy power-law scaling of the renormalized gap as a function of the bare one with an exponent which is interaction dependent.
\end{enumerate} 
This was revealed using general field theoretical arguments in the early eighties \cite{Kivelson85,Horovitz85,Wu86} and recently reinvestigated employing continuum field theories constructed as closely as possible to the microscopic model of interest \cite{gangadharaiah_etal_prl_12,RM_frg_prb_20,rational_boundary_charge_in_1d_prr_20}. Based on these insights one expects that the power-law renormalization of all band gaps is also found in microscopic models with more complex unit cells and $Z \geq 2$.   

 With this sanity check passed, in Ref.~\onlinecite{RM_frg_prb_20} we went beyond established interaction effects. We studied the decay of the inter unit cell density oscillations of the interacting RM model away from an open boundary for a Fermi energy placed within the gap. From this the associated boundary charge, accumulated close to the boundary, can be computed and was investigated. We finally computed the local spectral function. In Ref.~\cite{RM_frg_prb_20} as well as here we focus on the limit of vanishing temperature. We found 
 \begin{enumerate}[resume]
     \item that the density decay of the interacting gapped system remains dominantly exponential (as it is the case for vanishing interaction) but  
     that the behavior of the pre-exponential function is altered by the two-particle interaction and 
     \item that the characteristics (see below) of the fractional part of the boundary charge \cite{resta_prl_98,resta_sorella_prl_99,com,park_etal_prb_16,thakurathi_etal_prb_18,pletyukhov_etal_short,rational_boundary_charge_in_1d_prr_20,pletyukhov_kennes_klinovaja_loss_schoeller_prb_20,kingsmith_vanderbilt_prb_93} is robust against interaction effects and can fully be understood in terms of the (renormalized) bulk properties, while  
     \item the local spectral function close to the boundary might show an interaction induced in-gap $\delta$-peak associated to an effective edge-state which does not have a counterpart in the noninteracting limit.
 \end{enumerate}
It is remarkable that while the number of in-gap $\delta$ peaks of the spectral function, indicating effective edge states, depends on the two-particle interaction, the characteristics of the fractional part of the boundary charge is robust. This is of interest in the context of topological systems. It is known that in the noninteracting case topological bulk invariants are related to the number or parity of zero-energy edge states (bulk-boundary correspondence). These edge states are thus related to bulk properties. We will refer to them as conventional edge-states. In the interacting RM model the additional edge states instead originate from the local modulation of the self-energy close to the open boundary which goes beyond the intra unit cell structure. It follows from the interplay of the boundary and the two-particle interaction as can conveniently be illustrated using functional RG (see below). These edge states cannot be explained based on (renormalized) bulk properties and we refer to them as unconventional edge states.

The insights of Ref.~\cite{RM_frg_prb_20} raise the question if the properties 2.~to 4.~are specific to the interacting RM model or if they can be found in other, more complex lattice models of the above class as well---we already commented on the general expectation concerning the property 1.. To investigate this in a first step we here study the generalized Aubry-Andr\'e-Harper (AAH) model \cite{rational_boundary_charge_in_1d_prr_20,pletyukhov_kennes_klinovaja_loss_schoeller_prb_20,park_etal_prb_16,thakurathi_etal_prb_18,pletyukhov_etal_short,Harper_55,AA_80,Ganeshan_13,Lahini_09,Kraus_12,DeGottardi_13} with $Z=4$ and (homogeneous) nearest-neighbor interaction at different fillings $f$ and the Fermi energy falling in one of the $Z-1=3$ single-particle gaps. We confirm the properties 1.~to 4.~using functional RG. 

In a second step we pose another question: Is it conceivable to construct a model with a homogeneous single-particle part---thus being gapless in the noninteracting limit---which shows a similar phenomenology as described above but rooted in a periodically modulated, local two-particle interaction? Considering the cases of $Z=2$ and $Z=4$ periodicity and using functional RG as well as DMRG we show that this is indeed possible. It can be understood as follows. On the Hartree-Fock level the modulated two-particle interaction generates a modulated onsite energy and/or modulated effective hopping, and thus $Z-1$ (Hartree-Fock) gaps open. If the Fermi energy is chosen such that it lies inside one of these gaps the RG procedure, which contains the Hartree-Fock terms,
but other diagrammatic contributions on top, 
leads to the effects summarized in the above phenomenology. Within functional RG and for $Z=2$ this can even be shown analytically.

However,
there is an exception to this for $Z=4$. If one selects the Fermi energy such that it lies in the gap associated to half filling the gap opening is of second order in the interaction. As our approximate functional RG procedure does not contain all terms to second order we cannot use it to study this special case. Also DMRG is not the right tool to investigate the low-energy physics in this case. For small interactions the gap, setting the low-energy scale, is very small (of second order). It would thus require very large systems inaccessible by DMRG to study the half-filled situation. We thus leave the detailed study of this special case to future work.   

The modulated interaction model was studied earlier in Ref.~\onlinecite{ZW_Zuo_20}. Using DMRG and field theory the authors provided evidence that due to the modulation of the two-particle interaction a gap opens. However, their DMRG results for the density of a finite system with two open boundaries were not obtained in the ground state. The authors overlooked that with this method it is difficult to discriminate between states which are almost degenerate, a situation naturally arising in the presence of a zero-energy edge state (see below). In that case the density the authors present is the one of a linear combination of the ground and the first excited state.

Combining the results of the present paper with the ones of Ref.~\onlinecite{RM_frg_prb_20} provides strong evidence that the phenomenological properties 1.~to 4.~hold quite generally for 1D interacting lattice models with periodically modulated single-particle parameters and, under certain conditions, similar effects can also be found if the underlying noninteracting model has homogeneous parameters but the local interaction is periodically modulated around a finite average value.

We are not aware of any experimental realization of the 1d spinless modulated interaction model so far. However, first steps towards the experimental realization of extended, spinful models such as the Fermi-Hubbard model were taken. One of the candidates is a semiconductor heterostructure leading to a chain of gate-defined quantum dots, see Ref.~\cite{Hensgens_10}. Another possibility would be a cold Fermi gas in an optical lattice. In the bosonic case a modulated interaction model was e.g. realized in Ref.~\cite{Yamazaki_10}. For possibilities to realize noninteracting AAH models in various systems we refer to Refs.~\cite{Lahini_09,schreiber_etal_science_15,Ganeshan_13}

The rest of this paper is organized as follows. In the next section we present the models we consider, give a brief account of the many-body methods employed, and introduce the observables we study. Our results for the interacting generalized AAH model are presented in Sect.~\ref{sec:AAH}. 
Section \ref{sec:U_mod} is devoted to the results obtained for the models with periodically modulated interaction ($Z=2$ and $Z=4$). In Sect.~\ref{sec:summary} we summarize our results.  The functional RG flow equations are presented in Appendix \ref{App_a} while Appendix \ref{App_b} contains the details of the analytical solution of the flow equations for the modulated $U$ model with $Z=2$. In Appendix \ref{App_c} we briefly discuss the model with $Z=4$ at half filling and periodically modulated interaction.

We emphasize, that the present work should be viewed as a follow-up of Ref.~\onlinecite{RM_frg_prb_20}, while still being self contained to an appropriate degree. We thus refrain from presenting all the technical details and a comprehensive account of earlier works; we here restrict ourselves to the ones which are of direct relevance for our work. We furthermore only give a brief summary of the behavior of the observables, i.e., the local single-particle spectral function, the local density, and the boundary charge, in the noninteracting limit; our focus is on the changes due to the two-particle interaction.  For further details we refer the reader to Refs.~\onlinecite{RM_frg_prb_20}, \onlinecite{rational_boundary_charge_in_1d_prr_20,pletyukhov_kennes_klinovaja_loss_schoeller_prb_20}. 

\section{Models, methods, and observables}
\label{sec:modelphys}

\subsection{Models}
\label{sec:model}

\subsubsection{The interacting generalized Aubry-Andr\'e-Harper model with $Z=4$}

The first model we consider is the generalized 1D spinless AAH model with uniform nearest-neighbor two-particle interaction. The noninteracting model is given by the Hamiltonian
\begin{align}
  &H^{Z}_{\rm AAH} = \sum_j  \left( V_j n^{\phantom{}}_{j} 
  -  \left[t_j  c_{j+1}^\dag c_{j}^{\phantom{}} + \mbox{H.c.} \right] \right), \label{eq:H_AAH} 
  \end{align} 
  where $c_{j}^{(\dag)}$ is the second quantized annihilation (creation) operator at the site $j$ and $n_j=c_{j}^\dag c_{j}^{\phantom{}}$ the local density operator. The on-site potential $V_j$ and the hopping parameters $t_j$ are periodic with period $Z$, defining the number of lattice sites of the unit cell, 
  \begin{align}   &V_j=V\cos \left(\frac{2\pi j}{Z}+\varphi_v\right),
    \label{eq:V}\\
    &t_j = t +\delta t \cos\left(\frac{2\pi j}{Z}+\varphi_t
    \right).
  \label{eq:t} 
  \end{align}
 Here $V$ and $\delta t$ denote the amplitude of the modulation of the on-site potential and hopping, respectively. Furthermore, $\varphi_v $ and $\varphi_t$ are the phases of the corresponding modulation.  For generic parameters the noninteracting model has $Z-1$ band gaps of size $2 \Delta_\nu$, with $\nu=1,2,\ldots,Z-1$ \cite{pletyukhov_kennes_klinovaja_loss_schoeller_prb_20}.  
  
  We take $t$ as our unit of energy and set $t=1$. Moreover, we use $W=2t$ to denote half of the bandwidth of the (gapless) model with $\delta t=V=0$.

  The homogeneous two-particle interaction is assumed to be of nearest neighbor form and given by
  \begin{align}
	H_{\rm int} = U \sum_j \left( n^{\phantom{}}_{j} -\frac{1}{2} \right)
  \left( n^{\phantom{}}_{j+1} -\frac{1}{2} \right)  ,
 \label{eq:Hint}  
\end{align} with the (repulsive) interaction of strength $U \geq 0$. 

We take the total number of lattice sites $N$ to be an integer multiple of the period $Z$ such that all of the unit cells remain intact. One can rewrite the index $j$ of the Wannier basis into a unit cell index $n$ and intra-cell site index $i$
\begin{align}
  j=Z(n-1)+i.
\end{align}

To study the bulk properties of the system we consider periodic boundary conditions (PBCs). In this case the site index $j$ in the sum of equations (\ref{eq:H_AAH}) and (\ref{eq:Hint}) runs from $1$ to $N$ and we identify sites $N+1$ and $1$. To study the boundary physics, we take open boundary conditions (OBCs). In this case the sum in the above equations runs from $1$ to $N$ for the diagonal addends and from $1$ to $N-1$ for the off-diagonal ones. 

For $Z=2$, Eq.~(\ref{eq:H_AAH}) is the well-known Rice-Mele model, which is one of the basic models considered in the field of topological Fermi systems. In particular, it shows edge-state physics. In an earlier publication \cite{RM_frg_prb_20} we have provided a comprehensive study of interaction effects in this model. Our focus was on directly tackling the microscopic lattice model using the functional RG and DMRG. In addition, we used the mapping to field theories and methods applicable to those such as bosonization.   
As an extension beyond $Z=2$ we consider the interacting generalized AAH model with period $Z=4$ in the present paper. However, we already now note that our methods are applicable for arbitrary $Z$. 

\subsubsection{A tight-binding model with periodically modulated interaction}

In a second extension of our RM model study we investigate a tight-binding model with uniform hopping and vanishing single-particle potential but a periodically  modulated nearest-neighbor interaction.
We thus leave the class of models in which the gap, the edge-state physics, and the topological properties are present already in the noninteracting limit. 
Throughout the paper we consider interactions sufficiently small such that the homogeneous part does not lead to the opening of a correlation-induced gap (as it happens for our interaction Eq.~(\ref{eq:Hint}) at half filling for sufficiently large repulsive interactions \cite{Giamarchi03}).  
If any of the above phenomena 1.~to 4.~is found for the modulated interaction model, it thus originates from the inhomogeneous (but periodic) part of the two-particle interaction. The noninteracting Hamiltonian is 
\begin{align}
  &H_0 = - t \sum_j  \left( 
      c_{j+1}^\dag c_{j}^{\phantom{}} + \mbox{h.c.} \right) . \label{eq:H0} 
\end{align}
The interaction is given by
\begin{align}
  &H^{\rm Z}_{\rm int} =  \sum_j U_j \left( n^{\phantom{}}_{j} -\frac{1}{2} \right)
  \left( n^{\phantom{}}_{j+1} -\frac{1}{2} \right), 
  \label{eq:Hintinhom}
\end{align}
with
\begin{align}
  &U_{j}=U+\delta U \cos\left(\frac{2 \pi j}{Z}+\varphi_U\right)
  \label{eq:Umod}
\end{align}
where $\delta U$ is the amplitude of the modulation, $U \geq 0$ is the average value, and $\varphi_U$ is the phase. We, in particular, consider $Z=2$, with $U_j=U_{j+2}$ and $Z=4$ with $U_j=U_{j+4}$ for all $j$.
The same statements on the boundary conditions and the related summation limits as made for the generalized AAH model hold. 

\subsection{Many-body methods}
\label{sec:methods}

To solve the interacting models we employ the same methods as used in our interacting RM model study \cite{RM_frg_prb_20}, i.e., functional RG \cite{Metzner12} and DMRG \cite{Schollwoeck11}. In addition, we refer to field theoretical considerations employed to effective low-energy models \cite{Kivelson85,Horovitz85,Wu86,Jackiw83,Jackiw76,gangadharaiah_etal_prl_12,rational_boundary_charge_in_1d_prr_20}.  
To avoid any doubling we refer the reader interested in an introduction to the application of these methods to
(single-particle) gapped systems to  
Sect.~III of Ref.~\onlinecite{RM_frg_prb_20}. We here only summarize the individual 
advantages and shortcomings of the three approaches.

Functional RG and DMRG can directly be applied to the microscopic lattice models while the use of field theoretical tools requires the mapping to a continuum model. This only holds in the low-energy limit and for sufficiently weak two-particle interactions. However, within the field theory one can obtain analytical insights. 
In limiting cases this is also possible within functional RG.
As we use the lowest-order truncated approach the functional RG results are only controlled for small to intermediate two-particle interactions. If properly executed, DMRG can be considered as numerically exact. When being interested in boundary effects, we have to employ finite size DMRG. Due to the growth of the entanglement entropy it is, however, limited to system sizes of the order of $10^3$ lattice sites. The inverse system size sets a lower bound for the smallest accessible energy. In contrast, functional RG can be applied for very large systems allowing to access the asymptotic low-energy limit. Importantly, the lowest-order truncated functional RG was shown to be capable to capture the entire series of leading logarithms for ungapped \cite{Andergassen04,Karrasch10} as well as (single-particle) gapped \cite{RM_frg_prb_20} many fermion systems. This was crucial for the RM model \cite{RM_frg_prb_20} and will turn out to be equally important in our present extensions. 

To derive a closed set of functional RG flow equations for the self-energy and the local density of our models we employed the same approximations and cutoff procedure as discussed in Ref.~\onlinecite{RM_frg_prb_20}. Due to the enlarged size of the unit cell as compared to the $Z=2$ RM model studied in this paper, they become more complex. For completeness we present the RG equations in real space in Appendix \ref{App_a_real} and in momentum space in \ref{App_a_momentum}.  Both can easily be solved on a computer and in limiting cases even analytically; see Appendix \ref{App_b}. 

The functional RG approach truncated to lowest order in the two-particle interaction leads to an approximate self-energy which is frequency independent. At the end of the RG flow one thus has to deal with an effective Hamiltonian with single-particle parameters (hoppings and onsite energies) which are renormalized by the two particle interaction. For periodic boundary conditions the periodicity with period $Z$ is preserved. In the presence of open boundaries the effective single-particle parameters are modulated beyond the intra unit cell structure. This additional structure decays from the boundary towards the bulk and follows from the interplay of the boundary and the two-particle interaction. For large distances the renormalized bulk values obtained for periodic boundary conditions are reached. The approximate effective single-particle picture of the interacting problem helps to interpret the results (see Ref.~\cite{RM_frg_prb_20} and below).    

To obtain the DMRG data for the spectral gap and the local density we used the same approach and benchmarking as described in Sect. III.C of Ref.~\onlinecite{RM_frg_prb_20}. 

In Ref.~\onlinecite{RM_frg_prb_20} we have shown that to capture the physics
of the microscopic lattice model beyond leading order (in the two-particle interaction) it is advantageous to
construct a field theory directly from the model under consideration and use additional information available such as, e.g., the exact Bethe ansatz solution of the interacting homogeneous model. Here we do not aim at this level of accuracy and instead use general field theoretical arguments which do not contain any information on the underlying microscopic model beyond the leading order ones. For our present purposes this is sufficient. 

\subsection{Observables}
\label{sec:Observable}

We briefly introduce the observables we compute to investigate the effects of the two-particle interaction. Besides the band gaps we study the local particle density, the boundary charge, and the single-particle spectral function. 

We note that for our purposes the details of the analytical results which can be obtained for the noninteracting $Z=4$ generalized AAH model are not crucial. We thus refer the interested reader to Ref.~\cite{pletyukhov_kennes_klinovaja_loss_schoeller_prb_20} for a discussion of those.

\subsubsection{The gap}

In the noninteracting limit the three single-particle band gaps of size $2 \Delta_\nu$, with index $\nu=1,2,3$, of the $Z=4$ generalized AAH model are determined by the single-particle parameters $V$, $\delta t$, $\varphi_v$, and $\varphi_t$. Here we are not interested in the details of this dependence \cite{rational_boundary_charge_in_1d_prr_20,pletyukhov_kennes_klinovaja_loss_schoeller_prb_20} but rather on how the gap is modified by the two-particle interaction if the Fermi energy is placed in one of the gaps. From field-theoretical considerations \cite{Kivelson85,Horovitz85,Wu86,Jackiw83,Jackiw76,gangadharaiah_etal_prl_12,rational_boundary_charge_in_1d_prr_20}, which are independent of the details of the underlying lattice model, and our study of the $Z=2$ RM model \cite{RM_frg_prb_20} we expect power-law scaling of the renormalized gap as a function of the bare one. The exponent will depend on $U$. In addition, we expect a dependence on the band filling $f=\nu/Z$ associated to the corresponding index $\nu$
via the dependence on the Fermi momentum $k_F=\pi f$ where the gap opens; quarter filling for $\nu=1$, half filling for $\nu=2$, and three-quarter filling for $\nu=3$.   

For the modulated interaction model the noninteracting limit is gapless. However, the periodic modulation of the two-particle interaction might lead to gaps. This should not be confused with the gap which at half-filling opens even for homogeneous interactions if $U$ becomes sufficiently large \cite{Giamarchi03}. 
To study the possibility of a gap opening by DMRG we consider the difference of the energy of the first excited state and the ground state energy for systems of up to ${\mathcal O}(10^3)$ sites. Within lowest-order truncated functional RG we can employ the effective single-particle picture. If the modulated interaction will lead to modulated effective single-particle parameters, a gap might open. We can study this using the noninteracting formulas introducing the renormalized parameters.
The same can be done for the $Z=4$ interacting AAH model. Significantly larger systems than in DMRG or even the thermodynamic limit can be studied within functional RG (see below). 

\subsubsection{The local spectral function}
\label{sec:Observable_BSF}

The single-particle spectral function $A_j(\omega)$ is of particular interest if we consider a system with  open boundaries. It, on the one hand, shows the distribution of spectral weight $\left|\psi_{k}^{(\alpha)} (j)\right|^2$, given by the delocalized  single-particle wave functions  $\psi_{k}^{(\alpha)} (j)$ over the band energies $\varepsilon^{\alpha}_k$, with 
band index $\alpha$ and quasi-momentum $k$ for given site $j$. On the other hand possible edge states show up as in-gap $\delta$-peaks at their energy $\varepsilon_{\rm e}$ with a weight $\left| \psi_{\rm e} (j) \right|^2$ and wave function $ \psi_{\rm e} (j)$. In the noninteracting limit the eigenenergies and eigenstates can be obtained straightforwardly by numerically diagonalizing the Hamiltonian for a large but finite system (for analytical results, see Ref.~\cite{pletyukhov_kennes_klinovaja_loss_schoeller_prb_20}). The same can be done with the effective Hamiltonian at the end of the RG flow leading to an approximation of the local spectral function for $U >0$. This holds for the interacting $Z=4$ generalized AAH model as well as for the modulated interaction model at $Z=2$ and $Z=4$. 

To obtain the full spectral information from DMRG is computationally rather challenging and suffers from certain shortcomings \cite{J02,BSW09,KKM14}, mainly a limited energy resolution due to finite size effects, a finite bond dimension, and/or an artificial broadening. We thus refrain from using DMRG to obtain $A_j(\omega)$. 

In Ref.~\cite{RM_frg_prb_20} we have shown for the $Z=2$ RM model that besides the conventional edge-states additional interaction induced unconventional edge states appear in certain parts of the parameter space. They result from the spatial modulation of the renormalized single-particle parameters beyond the intra unit cell structure which decays from the boundaries towards the bulk. We investigate if the same happens in the $Z=4$ generalized AAH model. For the modulated interaction model both types of edge states, conventional and unconventional ones, might be induced by the interaction. In the former case the renormalized bulk parameters are such that the corresponding noninteracting model with these parameters shows an edge state. For the unconventional edge state this is not the case and they follow from the interplay of the boundary and the interaction as described above.

\subsubsection{The local density}
\label{sec:Observable_LPD}

For a noninteracting system it is straightforward to compute the local density $\rho(j)$ for a large but finite system by exact diagonalization (for analytical insights for the $Z=4$ generalized AAH model, see Ref.~\cite{pletyukhov_kennes_klinovaja_loss_schoeller_prb_20}). For models of the present type it can have two contributions. One coming from the filled bands $\rho_{\rm band}(j)$ and the other one coming from the filled edge states $\rho_{\rm edge}(j)$. In the presence of an open boundary the envelope modulating the intra unit cell density structure of $\rho_{\rm band}(j)$  from the boundary towards the bulk decays exponentially if the Fermi energy lies in one of the gaps. The decay length is given by the single-particle parameters (for analytical results for the $Z=4$ generalized AAH model, see Ref.~\cite{pletyukhov_kennes_klinovaja_loss_schoeller_prb_20}). The same holds for $\rho_{\rm edge}(j)$. 

To obtain the density of interacting systems by functional RG we used two approaches. One is to employ the effective single-particle picture and diagonalize the effective Hamiltonian at the end of the RG flow. However, it was earlier shown for metallic \cite{Andergassen04} as well as single-particle gapped models \cite{RM_frg_prb_20} that writing down a flow equation for the density leads to improved results. The corresponding flow equation is presented in Appendix  \ref{App_a}. We therefore also here rely on this approach which can be used for systems of up to $10^6$ lattice sites. 

Within DMRG the ground state density is directly accessible and we can thus use the numerically exact results for systems of the order of $10^3$ sites to compare to the approximate functional RG ones. 

For the $Z=2$ interacting RM model we have shown that the decay of the envelope of the density remains exponential with a renormalized decay length but that the pre-exponential function is altered by the interaction. We investigate if the same holds for the interacting $Z=4$ generalized AAH model.

For the modulated interaction models the noninteracting density away from the boundary decays generically as $1/j$. This is the decay of ordinary Friedel oscillations of metallic systems in one spatial dimension \cite{Mahan00}. However, at half-filling the particle-hole symmetry prohibits Friedel oscillations and the local density even in the presence of open boundaries is homogeneous \cite{Andergassen04} (and equal to $1/2$). This also holds in the presence of the nearest-neighbor interaction. We thus investigate the density of the $Z=4$ modulated interaction model with the Fermi energy placed in the gaps with indices $\nu=1,3$ corresponding to one- and three-quarter filling. 

\subsubsection{The boundary charge}
\label{sec:Observable_BC}
In the presence of an open boundary, charge might be accumulated close to it. This boundary charge $Q_{\rm B}$ can be computed as 
\begin{align}
    Q_{\rm B} = \lim_{M\rightarrow\infty}\lim_{N\rightarrow\infty}\sum_{j=1}^\infty \left[ \rho(j)-\bar{\rho}\right] f_{N,M}(j),
    \label{eq:QB_def}
\end{align} 
where the bulk averaged particle density is given as
\begin{align}
    \bar{\rho}=\frac{1}{Z}\sum^{Z}_{i=1}\rho_{\text{bulk}}(i) ,
\label{eq:Q_bulkave_def}
\end{align} 
with $\rho_{\text{bulk}}(i)$ computed for periodic boundary conditions, 
and $f_{N,M}=1-\theta_{MZ}(j-NZ)$ is an envelope function which defines the range of the boundary on the scale $NZ$ and varies smoothly from unity to zero on the scale $MZ$ (see also Fig.~3 of Ref.~\cite{rational_boundary_charge_in_1d_prr_20} for a sketch).
 Here, $\theta_{\delta x}(x)$ denotes some representation of the $\theta$-function with broadening $\delta x$. 
One needs to take $N\gg M\gg Z$ for the boundary charge to become independent of $M$ and $N$. As described in the last subsection $\rho(j)$ as well as $\rho_{\rm bulk}(i)$ are accessible by functional RG and DMRG. 

One of the most important universal properties of $Q_{\rm B}$ discussed in all detail in Refs.~\cite{park_etal_prb_16,thakurathi_etal_prb_18,pletyukhov_etal_short,pletyukhov_kennes_klinovaja_loss_schoeller_prb_20,rational_boundary_charge_in_1d_prr_20,weber_etal_prl_21} is its transformation under a shift of the lattice by one site towards the boundary, described by a change $\varphi\rightarrow\varphi + 2\pi/Z$ of all modulation phases $\varphi_v$, $\varphi_t$, and $\varphi_U$. For generalized AAH models it was shown for the noninteracting case in Refs.~\cite{pletyukhov_etal_short,pletyukhov_kennes_klinovaja_loss_schoeller_prb_20} that the boundary charge can only change by the average particle charge $f$ or the hole charge $f-1$ moved into the boundary, with $f=\nu/Z$. Therefore, as function of $\varphi=\varphi_t=\varphi_v$, one expects one of the two following possibilities for the phase dependence of $Q_{\rm B}$
\begin{align}
    \label{eq:QB_generic}
    Q_{\rm B} = f(\varphi) + F(\varphi) + \frac{\varphi}{2\pi}\begin{cases} \nu \\ \nu-Z \end{cases} \,,
\end{align}
where $f(\varphi)=f(\varphi+2\pi/Z)$ is some smooth and periodic function, and $F(\varphi)$ contains discrete jumps by unity when edge states cross the Fermi energy. A similiar result is also expected in the presence of a modulated interaction (with $\varphi=\varphi_U$) since, as shown below, the gap is induced by an effective Hartree-Fock mechanism. For the generic case (including interactions, several channels, and random disorder) it is expected that $Q_{\rm B}$ will always change by $\nu/Z\,\text{mod}(1)$ when changing the phase by $2\pi/Z$. This was motivated in Ref.~\cite{rational_boundary_charge_in_1d_prr_20} based on the nearsightedness principle that charge correlations in insulators will decay exponentially fast. 

In addition, the universal properties of the boundary charge have also been discussed within low-energy field-theoretical models for the case of small gaps (including interactions via bosonization methods) \cite{rational_boundary_charge_in_1d_prr_20,weber_etal_prl_21}. In this case it was shown that the boundary charge can be written as
\begin{align}
    \label{eq:QB_low_energy}
    Q_{\rm B} = \frac{\gamma_\nu}{2\pi} + \frac{\nu}{2Z}\,,
\end{align}
where $\gamma_\nu$ denotes the phase of the gap parameter in gap $\nu$, resulting from the resonant processes connecting the two Fermi points $\pm k_F=\pm \pi f$. The relation of the phase $\gamma_\nu$ to the modulation phase $\varphi$ has been analysed in all detail in Ref.~\cite{rational_boundary_charge_in_1d_prr_20} and is in general quite non-trivial for half-filling $f=1/2$. Away from half-filling, one obtains
\begin{align}
    \label{eq:gap_phase_vs_modulation_phase}
    \gamma_\nu = \begin{cases} \nu \varphi + \text{const} & \text{for}\quad f < 1/2 \\
    (\nu-Z)\varphi + \text{const} & \text{for}\quad f > 1/2 \end{cases} \,.
\end{align}
Here, the constant part can often be fixed via Eq.~(\ref{eq:QB_low_energy}) by special symmetry points \cite{rational_boundary_charge_in_1d_prr_20}. The universal result  Eq.~(\ref{eq:QB_low_energy}) shows that the boundary charge is only sensitive to the phase variable $\gamma_\nu$ but not to the gap size in the low-energy regime. For $Z=2$, this has been confirmed numerically via DMRG and functional RG to hold in the interacting case as well \cite{RM_frg_prb_20,weber_etal_prl_21}.

\section{Results for the interacting generalized AAH  model with $Z=4$}
\label{sec:AAH}

We now present explicit results for the renormalized gap, the local spectral function close to the boundary, the local density, and the boundary charge for the interacting generalized AAH model with $Z=4$. As already emphasized, we are only interested in the cases in which the Fermi energy lies within one of the three gaps and the (noninteracting) system is a band insulator. Note that the Fermi energy depends on the two-particle interaction $U$ and must thus be chosen such that the target gap or the corresponding band filling is reached. There is a unique relation between the band gap index $\nu$ and the band filling $f$. 

From effective low-energy field theories we expect that for all $\nu$ the dependence of the renormalized gap  $\Delta^{\rm ren}_{\nu}$ on the bare one $\Delta_\nu$ is given by  
\begin{align}
    \frac{\Delta^{\text{ren}}_\nu}{\Delta_\nu}=\left(  \frac{\Delta_\nu}{W} \right)^{\beta_f(U)}  \text{ for } \Delta_\nu \ll W, 
    \label{eq:gap_generic_def}
\end{align} 
with an interaction and filling dependent exponent $\beta_f(U)$. 
Bosonization predicts that to leading order in the two-particle interaction $\beta_f(U)=K(U,f)-1$ with the Tomonaga-Luttinger liquid parameter $K$ of the underlying homogeneous model \cite{delft_schoeller_98,Giamarchi03,Schoenhammer05}. For the present case $K(U,f)-1=-U[1-2\cos (2 k_{\rm F})]/[2\pi\sin(k_{\rm F})] + {\mathcal O}(U^2)$ and $k_{\rm F}=f \pi$ \cite{Andergassen04,delft_schoeller_98,Giamarchi03,Schoenhammer05}.

We have earlier confirmed this for the microscopic $Z=2$ RM model at half-filling (the single gap for $Z=2$ corresponds to half-filling) directly without the approximate mapping to a field theory \cite{RM_frg_prb_20}.  

\begin{figure}[t]
   \centering
   \includegraphics[width=0.5\textwidth]{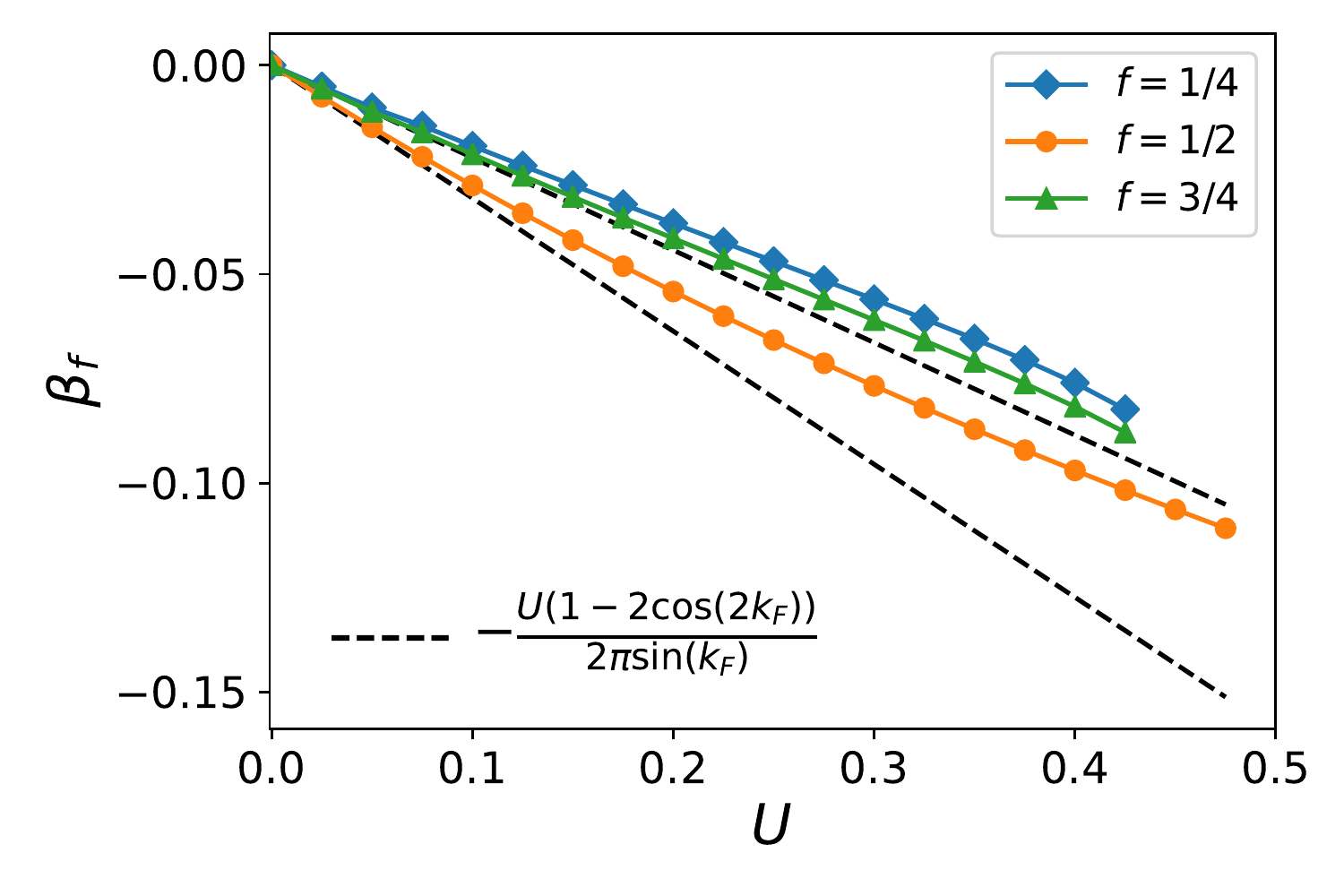}
   \\[0mm]
   \caption{The $U$-dependence of the exponent of the power-law scaling of the effective gap as a function of the bare one taken from functional RG data of the interacting generalized AAH model with $Z=4$. Here $f$ is the filling. The black dashed lines show the leading order exponent according to field theory. 
   Note that one-quarter and three-quarter filling lead to the same leading order exponent. 
   The lower dashed line is for $f=1/2$ the upper one for $f=1/4$ and $3/4$.
   The phases of the modulated single-particle parameters are $\varphi_v=\varphi_t=-0.45-0.25\pi$. 
       For quarter filling, the other parameters are $V=0.15,\delta t=0.2$, for half filling, $V=0.15,\delta t=0.075$, and for three-quarter filling, $V=0.15,\delta t=0.1$.
   }
   \label{fig:eff_gap_aah}
\end{figure}

Figure \ref{fig:eff_gap_aah} shows the exponent $\beta_f$ as a function of $U$ for the generalized interacting $Z=4$ AAH model at quarter, half, and three-quarter filling. The values of the single-particle parameters are given in the caption. For fixed $U$ and filling the exponent was determined as follows [see Eq.~(37) of Ref.~\onlinecite{RM_frg_prb_20}]: $\Delta_\nu^{\text{ren}}$ was computed from numerical diagonalization of the effective single-particle Hamiltonian in momentum space (thermodynamic limit) at the end of the functional RG flow (see Appendix \ref{App_a_momentum}). After dividing by the bare gap $\Delta_\nu$ logarithmic centered differences with respect to $\Delta_\nu$ were taken leading to a $U$ and filling dependent constant in the limit of small $\Delta_\nu$. This value corresponds to the exponent. In Fig.~\ref{fig:eff_gap_aah} the numerical data (symbols) are compared to the leading order in $U$ prediction from field theory shown as black dashed lines. Note that quarter and three-quarter fillings lead to the same leading order expression for $\beta_f$. The data agree to the field theoretical prediction to leading order in $U$. All this is as expected. To save computational resources we therefore refrain from presenting DMRG data for the gap scaling  of the interacting generalized $Z=4$ AAH model. 
How quickly the functional RG results for the exponent deviate from field theory obviously when increasing $U$ depends on the filling (gap index). We, however, emphasize, that employing the lowest-order truncated functional RG we do not control the orders of the exponent beyond the leading one; for an extensive discussion of this for the interacting RM model see Ref.~\cite{RM_frg_prb_20}.

\begin{figure}[h]
   \centering
   \includegraphics[width=0.5\textwidth]{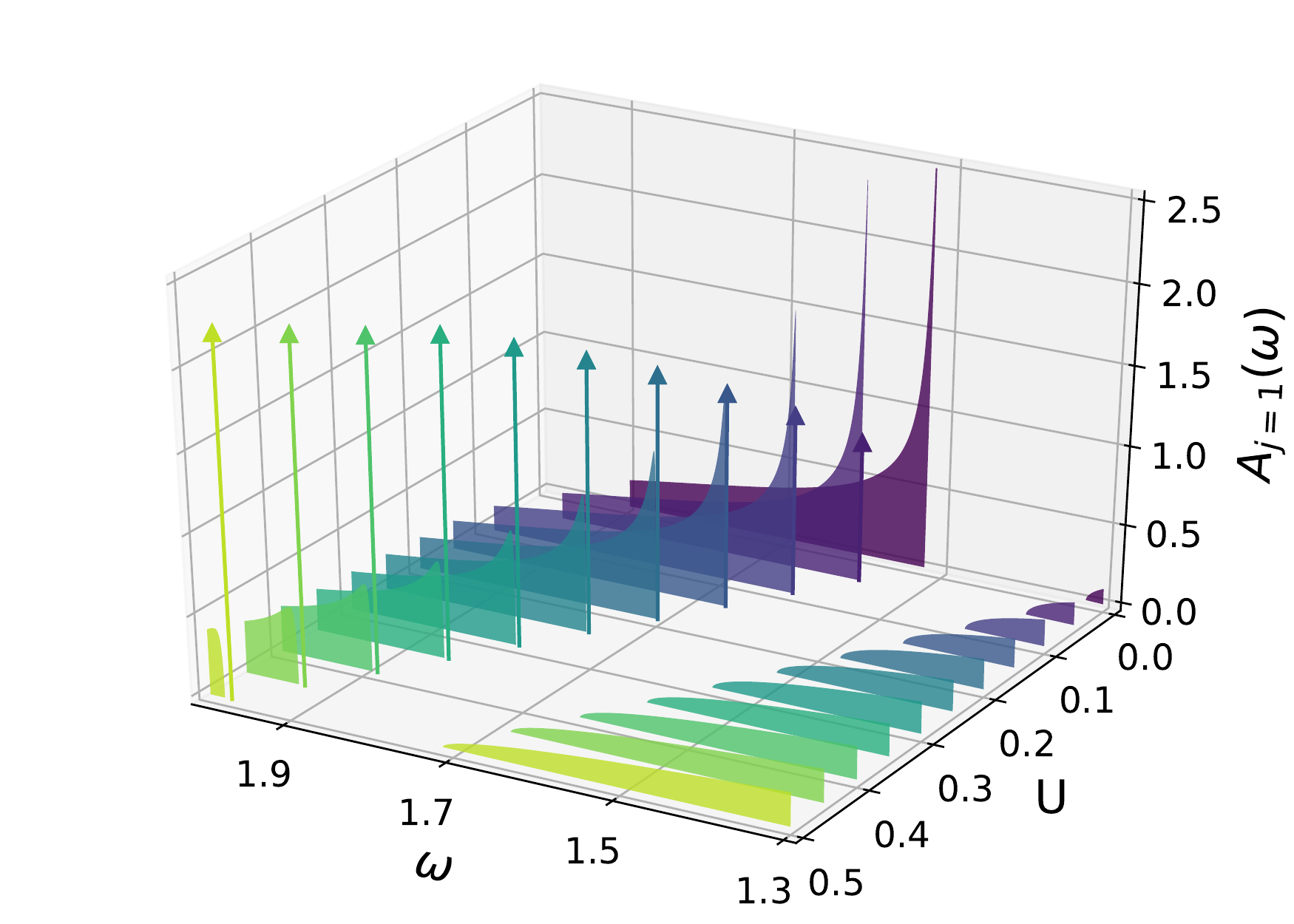}
   \\[0mm]
   \caption{Functional RG data of the local single-particle spectral function $A_j(\omega)$ of the interacting, generalized, $Z=4$ AAH model with open boundaries on site $j=1$ with $N=6000, V=0.27, \delta t=0.15, \varphi_v=-0.05,  \varphi_t=-0.17$. Data for three-quarter filling and different $U$ are shown. The interaction induced effective edge states are indicated by vertical arrows with the heights being proportional to the corresponding spectral weight.}
   \label{fig:lsf_aah_z4}
\end{figure}

In Fig.~\ref{fig:lsf_aah_z4} we show exemplary functional RG results for the local single-particle spectral function on site $j=1$ of the interacting generalized $Z=4$ AAH model with open boundaries. Three-quarter filling is chosen. Energies around the gap with index $\nu=3$ are shown corresponding to the low-energy regime (the Fermi energy lies in the center of the gap shown).  
The parameters, as given in the caption, are taken from a regime in which the $U=0$ function shows a van-Hove singularity at the band edge of the gap but no in-gap edge state (see the dark purple curve in Fig.~\ref{fig:lsf_aah_z4}). For the $Z=2$ RM model we have shown that this is the parameter regime in which with increasing $U$ an effective interaction-induced unconventional in-gap edge state forms \cite{RM_frg_prb_20}. As Fig.~\ref{fig:lsf_aah_z4} shows the same happens in the generalized AAH model. The edge state is indicated by an in-gap $\delta$-peak displayed by a vertical arrow of a height which is proportional to the spectral weight. With increasing $U$ the in-gap state detaches from the van-Hove singularity and gains weight. It is not located at the Fermi energy, i.e., it is not a zero energy edge state. The appearance of this interaction induced edge state can be understood as follows. During the RG flow the interplay of the open boundary and the interaction leads to the  build-up of a spatial modulation of the effective single-particle parameters beyond the unit cell structure. The corresponding envelope decays from the boundary towards the bulk and can lead to bound states located close to the boundary (edge states). Deep in the bulk the same effective single-particle parameters as obtained from a calculation with periodic boundary conditions is reached. Crucially, the appearance of such unconventional edge states cannot be understood from either the bare or the renormalized single-particle parameters in the bulk and does thus not follow the notion of the standard bulk-boundary correspondence which deals with conventional edge states only.  

The spectral function is computed for $N=6000$ lattice sites. To obtain a continuous function for energies within the bands the spectral weight is averaged over a few eigenenergies. Further increasing the system size does not lead to any visible changes on the scale of the plot and the results can be considered as being in the thermodynamic limit.  

As a side remark we emphasize that Fig.~\ref{fig:lsf_aah_z4} also illustrates that the gap increases with increasing $U$ (see above).

\begin{figure}[t]
  \begin{center}
  \includegraphics[width=0.5\textwidth,clip]{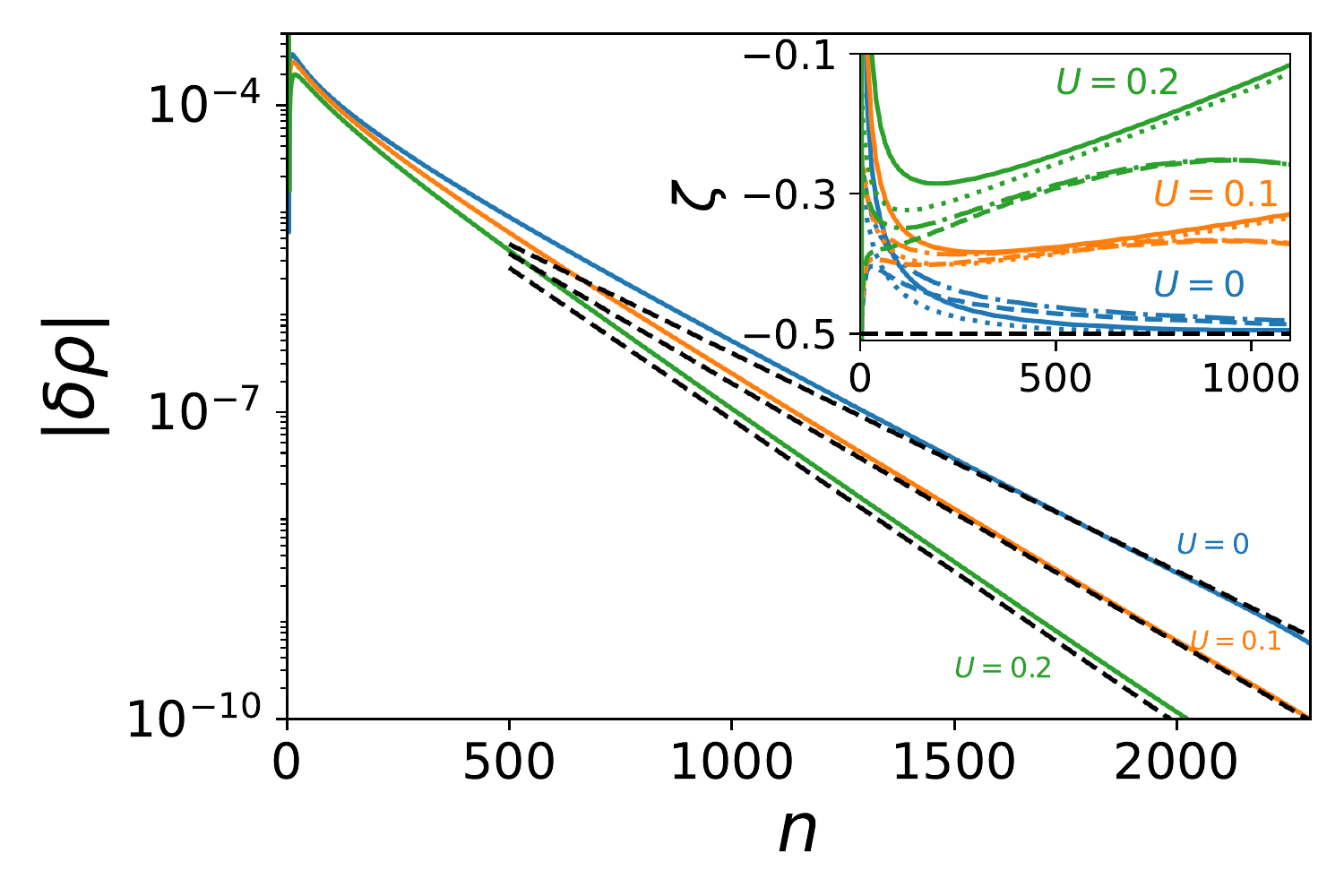}
  \caption{Main panel: Functional RG data for of the local density relative to the bulk value as a function of the unit cell index $n$ for $i=1$ and different $U$. The single-particle parameters are $N=20000$, $V=0.03$, $\delta t =0.08$, $\varphi_v=\varphi_t=\pi/4$.  Half filling is considered. Note the linear-log scale. The slope of the dashed line is computed plugging the renormalized single-particle parameters into the analytical expression for $-2\kappa^{\text{ren}}_{\text{bc}}$ at $U=0$, where $\kappa^{\text{ren}}_{\text{bc}}$ is an inverse decay length calculated in Ref.~\cite{pletyukhov_kennes_klinovaja_loss_schoeller_prb_20}. Inset: The logarithmic derivative of the pre-exponential function. Solid lines are for $i=1$, dashed lines are for $i=2$, dotted lines are for $i=3$ and the dashed-dotted ones are for $i=4$. }
  \label{fig:rho_AAH_halffilling}
  \end{center}
\end{figure}

The bulk density on each intra-cell lattice site $i$ takes a value which is independent of the unit cell index $n$. However, the density of the interacting generalized AAH model becomes $n$-dependent if a system with an open boundary is considered. The main part of Fig.~\ref{fig:rho_AAH_halffilling} shows exemplary functional RG results for the absolute value of the difference between the bulk density and the one obtained in the presence of an open boundary for the intra-cell index  $i=1$ and different $U$. To be able to access large distances from the boundary without any significant finite size effects a fairly large system size of $N=20000$ sites is considered. Such cannot be reached by DMRG.  Without loss of generality we focus on half filling. The single-particle parameters are given in the caption. For sufficiently large $n$ the decay towards the bulk follows an exponential function with a $U$-dependent decay length $(2\kappa_{\text {bp}})^{-1}$; note the linear-log scale of  Fig.~\ref{fig:rho_AAH_halffilling}. This length scale is independent of $i$ (not shown). For $U=0$ an analytic expression for $2\kappa_{\text {bp}}$ in terms of the single-particle parameters is known \cite{pletyukhov_kennes_klinovaja_loss_schoeller_prb_20}. The details of this formula are irrelevant in the present context and we do not reproduce it here. The figure indicates that this expression can also be used in the interacting case if the bulk renormalized single-particle parameters at the end of the RG flow are plugged into the noninteracting formula; see the dashed lines.  The decay length is linked to the gap size and it is thus not surprising, that $2\kappa_{\text {bp}}$ increases with increasing $U$.

After the exponential part of the $n$-dependence is known it can be divided out. For $U=0$ the resulting pre-exponential function decays as $n^{-1/2}$. This is shown in the inset of   Fig.~\ref{fig:rho_AAH_halffilling}, in which we present the log-derivative of the pre-exponential function as a function of $n$. For all $i$ the data approach the asymptotic exponent $\zeta=-1/2$. This is known analytically \cite{pletyukhov_kennes_klinovaja_loss_schoeller_prb_20}. For increasing interaction the pre-exponential function changes. In particular, it does no longer seem to decay as the inverse square root of the unit cell index but instead shows a richer behavior. As for the gap and the spectral function these findings are in full agreement with our results for the interacting RM model \cite{RM_frg_prb_20}. 

\begin{figure}[t]
  \begin{center}
  \includegraphics[width=0.5\textwidth,clip]{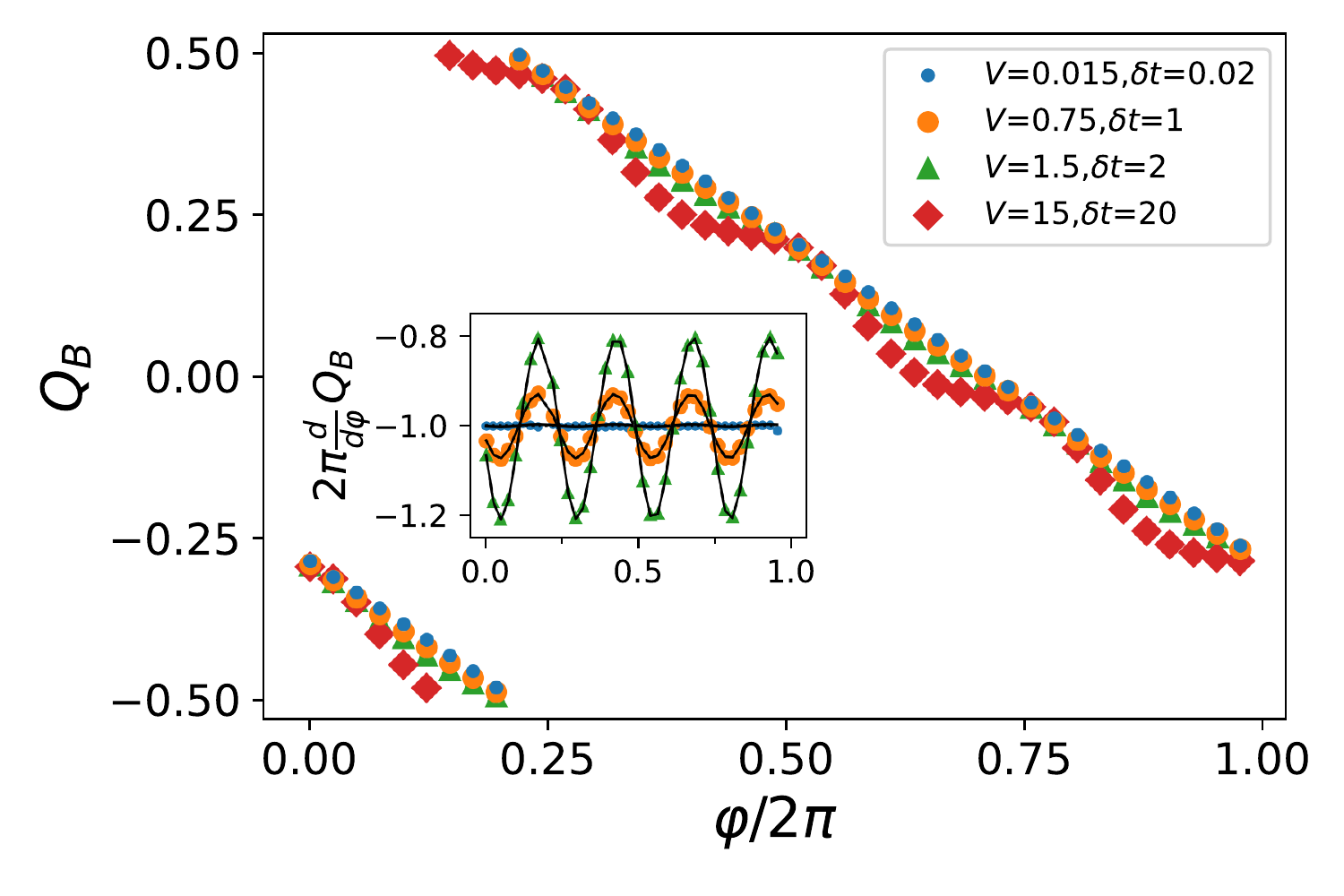}
  \caption{Main panel: Functional RG data for the boundary charge of the interacting three-quarter filled generalized AAH model as a function of $\varphi=\varphi_v=\varphi_t$ for different $\delta t$ and $V$. The system size is $N=4000$ and the interaction $U=0.08$. Inset: Derivative of the data of the main panel with respect to $\varphi$. Black lines are the result obtained from the noninteracting generalized AAH model but with the effective bulk renormalized parameters. } 
  \label{fig:bc_AAH_3qrtfilling}
  \end{center}
\end{figure}

Finally, we discuss the boundary charge of the interacting generalized AAH model. In Fig.~\ref{fig:bc_AAH_3qrtfilling} we show $Q_{\rm B}$ (symbols) as a function of $\varphi=\varphi_v=\varphi_t$ for $U=0.08$, different $V$ and $\delta t$ at three-quarter filling. 
The line shape is consistent with the second case of Eq.~(\ref{eq:QB_generic}) (involving $\nu-Z=-1$) and, for small gaps, with 
Eqs.~(\ref{eq:QB_low_energy}) and (\ref{eq:gap_phase_vs_modulation_phase}). The linear dependence of $Q_{\rm B}$ on $\varphi$ is known from noninteracting models \cite{park_etal_prb_16,thakurathi_etal_prb_18,pletyukhov_etal_short,pletyukhov_kennes_klinovaja_loss_schoeller_prb_20}, for small gaps from effective low-energy field theories \cite{rational_boundary_charge_in_1d_prr_20}, and from our results for the microscopic RM model \cite{RM_frg_prb_20}, see the summary in Sect.~\ref{sec:Observable_BC}. 
As already emphasized in these works it is robust for weak two-particle interactions. Moreover, the interaction enhanced corrections to the linear behavior can be understood from the effective single particle picture. In the inset of Fig.~\ref{fig:bc_AAH_3qrtfilling}, we show the derivative of the data of the corresponding main panel (symbols). Black lines are the results obtained from the  noninteracting AAH model but with the renormalized effective bulk parameters. They coincide with the numerical data. Therefore, the interacting boundary charge can be determined by the effective bulk parameters alone. 

For large gaps (red diamonds in Fig.~~\ref{fig:bc_AAH_3qrtfilling}) plateaus at multiples of $1/4$ develop. These are also known from considering the atomic limit \cite{rational_boundary_charge_in_1d_prr_20} and from our discussion of the microscopic RM model \cite{RM_frg_prb_20}. We found similar results for the one-quarter filled AAH model.

\begin{figure}[t]
  \begin{center}
  \includegraphics[width=0.5\textwidth,clip]{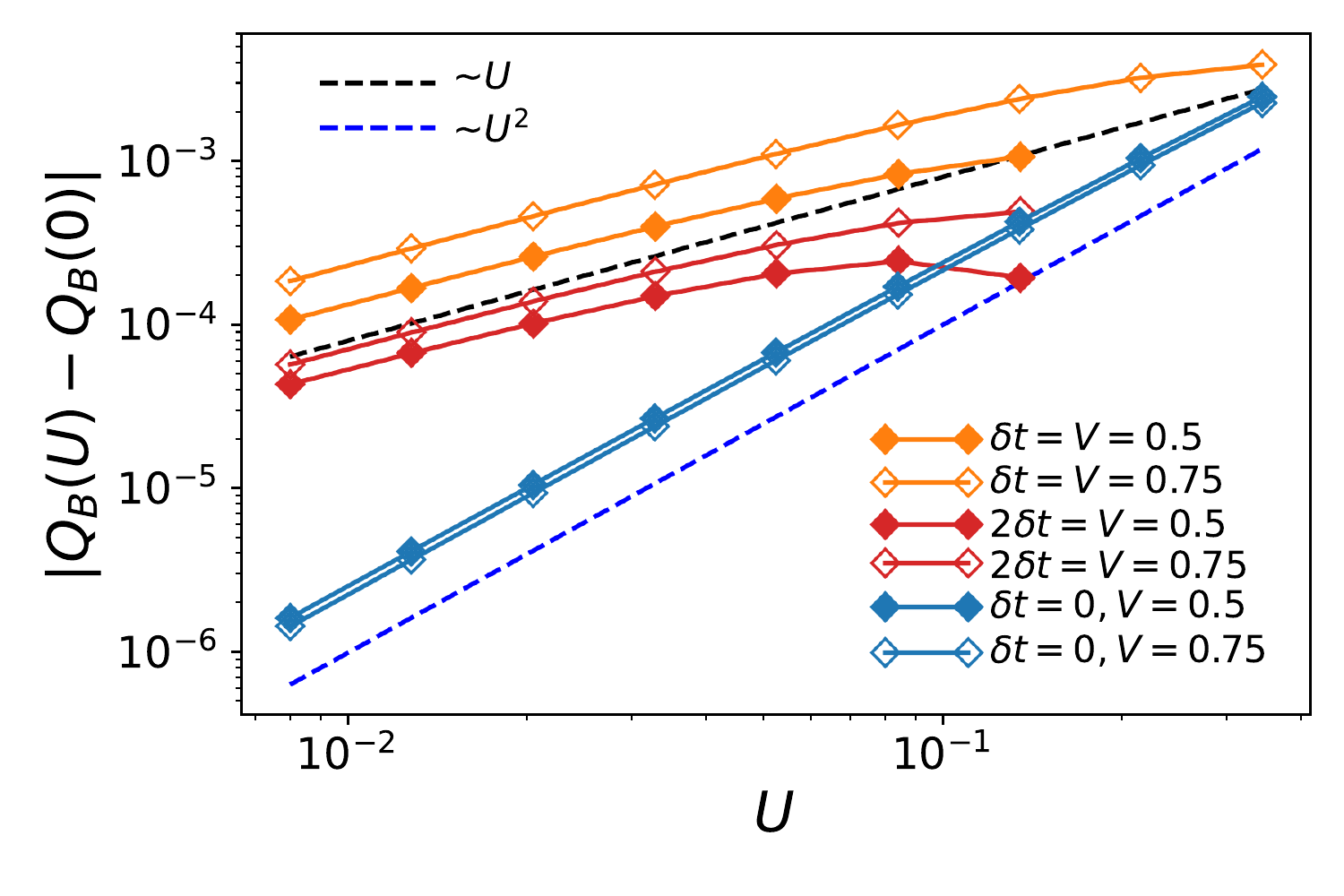}
  \caption{Functional RG data for the interaction correction of the boundary charge for the generalized AAH $Z=4$ model at half filling as a function of $U$. Different $V$ and $\delta t$ are considered and we choose $\varphi_v=\varphi_t=5\pi/6$. The system size is  $N=4000$. The black dashed line indicates a power law $\sim U$, and the blue dashed line a power-law $\sim U^2$. The weak $U$-dependence in the uniform hopping limit with $\delta t=0$ indicates that the quantization of the fractional part of the boundary charge is particularly robust against two-particle interactions.  Note the log-log scale. } 
  \label{fig:diff_qb_AAH_halffilling}
  \end{center}
\end{figure}

For the half-filled noninteracting generalized AAH model, there is no simple universal linear relation between the boundary charge and the phase of the modulation in the small gap limit. Instead, Ref.~\onlinecite{rational_boundary_charge_in_1d_prr_20} showed that the fractional part of the boundary charge is quantized in the uniform hopping limit $\delta t=0$  and takes values $\pm 1/2$ and $0$. In this work it was also shown, that the quantization is robust towards two-particle interactions (and disorder). In the following, we examine the robustness of the boundary charge for the half-filled interacting generalized AAH model using our functional RG approach. 
Away from the uniform hopping limit, the boundary charge at $U=0$ is not quantized and shows a nonuiversal behavior we here do not further investigate. We only studied the difference between the interacting and noninteracting boundary charge for both the unifom hopping limit and generic parameters to show that results for $Q_{\rm B}$ are in both cases rather robust towards the two-particle interaction.  
In Fig.~\ref{fig:diff_qb_AAH_halffilling} we show the absolute value of the interaction correction of the boundary charge for small interactions as a function of $U$ on a log-log scale. 
On the one hand generic exemplary single-particle parameters are taken (red and orange symbols), on the other hand the subspace with $\delta t=0$ (blue symbols) is considered. The black dashed line indicates a power law $\sim U$ and the blue one a power law $\sim U^2$. 
The detailed values of the single-particle parameters are given in the legend. 
 We observe that for all single-particle parameters the correction of the boundary charge is very small. Moreover, for the generic cases, the correction of the boundary charge scales linearly in $U$ in the small $U$ limit. 
In contrast, in the uniform hopping limit with $\delta t=0$, the linear $U$ correction vanishes rendering the quantized boundary charge particularly insensitive to the two-particle interaction. We note that our approximate functional RG procedure does not contain all terms of order $U^2$ thus we do not control the results in the uniform hopping limit besides the insight that the linear term vanishes. 

These results for the boundary charge again support our idea \cite{RM_frg_prb_20} that it might be more appropriate and a more physical indicator of the relation between boundary and bulk properties for interacting systems as compared to the number of edge states. 

The above insights on the interaction dependence of the gap, the single-particle spectral function, the local density, as well as the charge accumulated close to an open boundary of the generalized AAH model for $Z=4$ are in full accordance with the ones we gained for the $Z=2$ RM model. This indicates that the phenomenology on the interaction dependence of the above observables for gapped Fermi systems summarized by the properties 1. to 4. in the Introduction and first formulated for the RM model seems to be generalizable to models with more complex unit cell structure. 

We now proceed and investigate if effects similar to the ones summarized in this phenomenology are also found in models which are gapless in the noninteracting limit. The gap is instead generated by a spatially modulated interaction. 

\section{Results for the modulated interaction model}
\label{sec:U_mod}

\subsection{The $Z=2$ model}
\label{subsec:mod_u_Z=2}
	
In this section we discuss the modulated interaction model with $Z=2$. For symmetry reasons a gap can only open around zero band energy. In the following we thus consider a vanishing Fermi energy and half filling. Due to particle-hole symmetry no effective onsite potential is generated during the RG flow. As we have discussed in Ref.~\onlinecite{RM_frg_prb_20} this symmetry also implies that an open boundary does not induce any density modulations beyond the intra unit cell structure. For the present model we thus only consider the gap and the local spectral function as observables. 


To the best of our knowledge only the gap of the  modulated interaction model was so far studied in a reliable way \cite{ZW_Zuo_20}. We therefore have no expectations from other models or field theory we can compare and benchmark our approximate functional RG results for the ground state density and boundary charge against. For this reason numerically exact DMRG results obtainable for small to moderate system sizes will play a major role in this and the next subsection.


However, we first discuss approximate analytical insights which can be gained by functional RG. The RG flow equations can be found in Appendix \ref{App_a}. For periodic boundary conditions, the number of coupled flow equations is reduced to two: the ones for $t^{\Lambda}_1$ and $t^{\Lambda}_2$. This shows that an effective SSH model is generated by the modulation of the interaction. To the given approximation, the physics of this effective single-particle model corresponds to the one of the modulated interaction model. 

The effective flowing gap is determined by 
\begin{align}
    2\Delta^{\Lambda}=2\left| t^{\Lambda}_{\text{1}}-t^{\Lambda}_{\text{2}}\right|. \label{eq:effgap_u1u2} 
\end{align} 
In Appendix \ref{App_b}, we present the details of an analytical derivation of functional RG results for the effective gap in the small interaction and the small gap limit. In a first step we prove that the gap in first order perturbation theory for the self-energy, i.e.~on Hartree-Fock level, is given by 
\begin{align}
    2\Delta^{\text{HF}}=\frac{4 \delta U}{\pi}. \label{eq:HF_effgap_u1u2}
\end{align} 
In a second step we show that for $\Lambda_{\text{IR}}  \ll W$, the solution of the flow equation of the effective gap is given by
\begin{align}
	2\Delta(\Lambda_{\text{IR}})= 2\Delta^{\text{HF}}  \left(\frac{\Lambda_{\rm IR}}{4t}\right)^{-U/\pi t}(2-2^{U/\pi t}),	 \label{eq:gap_dU_scaling} 
	\end{align}
where $\Lambda_{\text{IR}}$ is the infrared (IR) scale, cutting off the RG flow. It needs to be determined numerically. In Fig.~\ref{fig:flow_gap_u1u2_smallU}, the momentum space RG flow for the effective gap divided by $\delta U$ is shown for very small $\delta U$. From this figure (and related ones for other parameter sets) we conclude in a third step that the IR cutoff is given by the Hartree-Fock gap $\Lambda^{\text{IR}}=2\Delta^{\text{HF}}$. This appears to be reasonable on general grounds and could have been guessed even without the numerical results.
The final result for the renormalized gap 
\begin{align}
	2\Delta^{\text{ren}}=  2\Delta^{\text{HF}} \left( \frac{\Delta^{\text{HF}}}{2 t} \right)^{-U/(\pi t)}(2-2^{U/\pi t})	 \label{eq:gap_dU_scaling_final} 
\end{align}
shows that in the RG procedure logarithms are resummed to a power law as it is the case in the interacting RM model and the interacting generalized AAH model.

We, however, emphasize that the appearance of the interaction in the basis [in form of the modulation amplitude $\delta U$; see Eq.~(\ref{eq:HF_effgap_u1u2})] and the exponent (in the form of the unit cell average $U$) makes it less obvious that our approximate functional RG  indeed captures the physics for small to moderate two-particle interactions.  

\begin{figure}[t]
   \centering
   \includegraphics[width=0.5\textwidth]{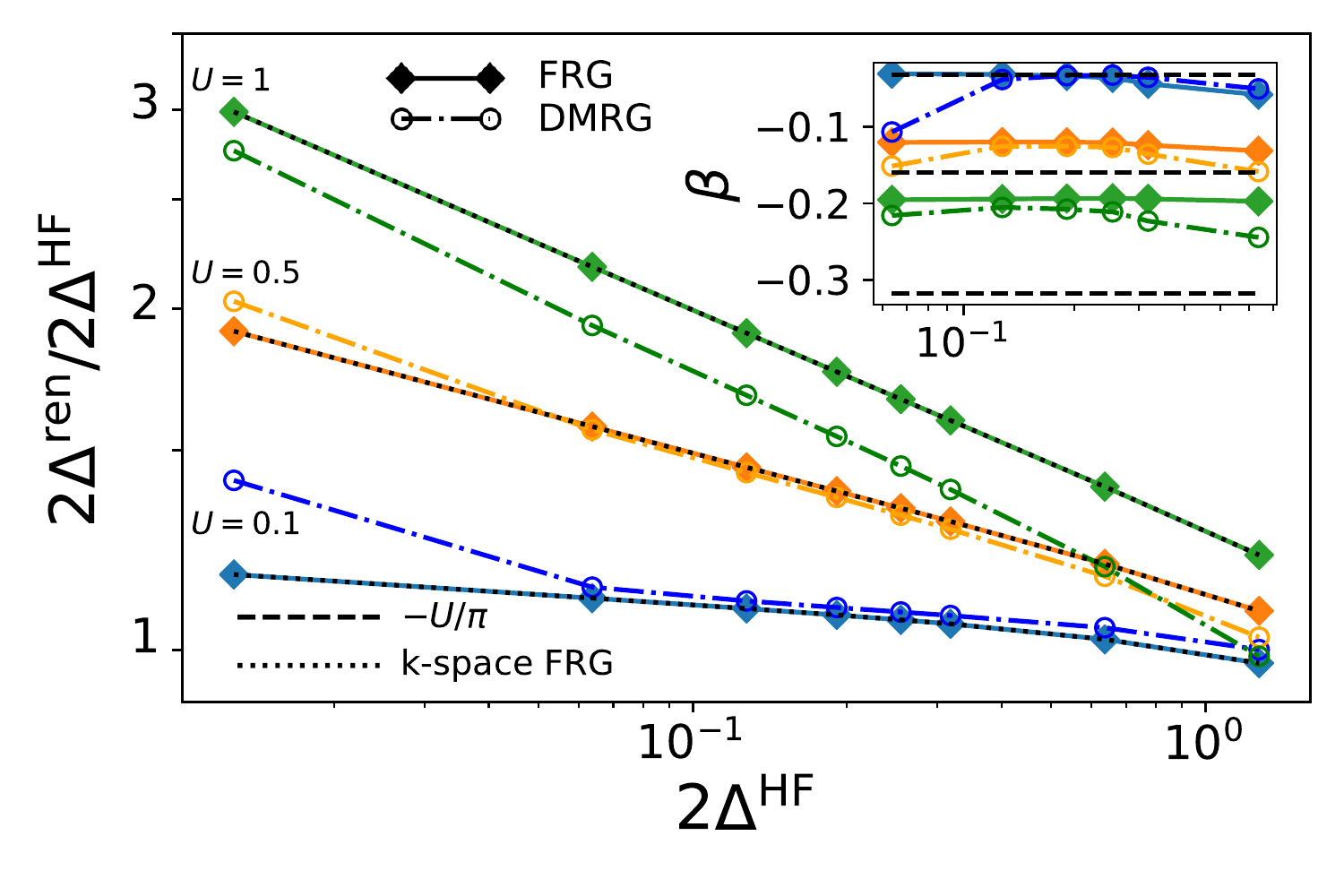}
   \\[0mm]
   \caption{Main plot: Ratio of the renormalized gap and the Hartree-Fock gap $2\Delta^{\mathrm{HF}}$ as a function of $2\Delta^{\mathrm{HF}}$.  Data of DMRG (open symbols), real-space FRG (solid lines with filled diamonds), and $k$-space functional RG (black dotted lines) are shown. Inset: The log-derivative and thus the apparent exponent $\beta$ of the data in the main plot. The system size is $N=1000$. The corresponding $\delta U$ are $0.01, 0.05, 0.1, 0.15, 0.2, 0.25, 0.5, 1.0$. Three different $U$ are considered. Note the log-log scale of the main plot.}
   \label{fig:FRGvsDMRG_gap_u1u2}
\end{figure}

In Fig.~\ref{fig:FRGvsDMRG_gap_u1u2}
we therefore compare numerical results for the gap at the end of the RG flow  obtained by  real-space as well as $k$-space functional RG with DMRG results.
We note that besides the lowest order trunction the numerical solution of the functional RG equations does not contain any further approximations.
This has to be contrasted to the analytical functional RG analysis of Appendix \ref{App_b} leading to Eq.~(\ref{eq:gap_dU_scaling_final}) which requires such steps. The main panel shows the comparison of the renormalized gap divided by $2\Delta^{\text{HF}}$ as a function of the Hartree-Fock gap. The real-space functional RG data are shown as color-coded filled diamonds. In comparison DMRG data are indicated by open symbols, and the momentum space  functional RG results are displayed as  dotted lines. 

For not too large interactions, that is $U=0.1$ and $U=0.5$, the functional RG data and the DMRG data agree nicely when $2\Delta^{\text{HF}}$ is not too small. The upturn of the DMRG data at small Hartree-Fock gaps and the resulting deviation from the functional RG data can be understood as follows. For very small gaps, very low energy scales must be accessible. In DMRG the low-energy scale is set by the inverse system size. If the gap becomes smaller than this scale, finite size effects prevail and in the present case lead to the upturn. Due to limitations in computing resources it is not possible to produce DMRG data which are converged with respect to the bond dimension---as it is the case in Fig.~\ref{fig:FRGvsDMRG_gap_u1u2}---for systems significantly larger than the $N=1000$ lattice sites considered here. The $k$-space functional RG results were obtained in the thermodynamic limit instead and do not suffer from finite size effects. The real-space functional RG data were, however, obtained for the same system size as studied by DMRG. Thus finite size effects matter for these as well but apparently do not show up in Fig.~\ref{fig:FRGvsDMRG_gap_u1u2}.  Two reasons for this appear to be reasonable. Firstly, the way the gap is extracted in finite size DMRG and real-space functional RG (gap in the many-body spectrum versus gap in the effective single-particle spectrum) differ. Secondly, truncated functional RG is an approximate tool, while DMRG (for finite systems) is numerically exact. Both might affect the details of the finite size corrections.

As the renormalized gap increases with increasing $U$ and thus the low-energy scale increases, this finite size effect becomes less severe for larger $U$. Accordingly, the $U=1$ DMRG data show barely any upturn at small $2\Delta^{\text{HF}}$. However, for this interaction higher-order corrections missed in the functional RG approach become important and DMRG and functional RG data deviate for all   $2\Delta^{\text{HF}}$ (see the green data).

Most notably all data sets of Fig.~\ref{fig:FRGvsDMRG_gap_u1u2} show linear behavior for intermediate $2\Delta^{\text{HF}}$ on a log-log scale. In the inset the log-derivative of the data is shown. Up to the finite-size issues just discussed and higher order (in $U$) corrections DMRG confirms the functional RG result Eq.~(\ref{eq:gap_dU_scaling_final}) of power-law scaling. We observe that the DMRG and functional RG exponents deviate from the leading order expression $-U/\pi$ (black dashed lines in the inset of Fig.~\ref{fig:FRGvsDMRG_gap_u1u2}) in similar ways. We have already observed this for the $Z=2$ interacting RM model. For a discussion of this, see Ref.~\onlinecite{RM_frg_prb_20}.  As the full numerical solution of the RG equations contains terms beyond the leading order, the deviation of the numerical exponent for $U$ of order 1 from the analytical exponent $-U/\pi$ is not surprising. 

All this confirms that our functional RG analysis based on the lowest-order truncation indeed provides the correct answer. A gap opens in the modulated interaction model which is triggered by the modulation amplitude $\delta U$ and scales as a power-law in the Hartree-Fock gap, with the interaction $U$ averaged over the unit cell entering the exponent to leading order. The leading order exponent $-U/\pi$ is the same as found in the $Z=2$ half-filled interacting RM model \cite{RM_frg_prb_20} as well as the $Z=4$ half-filled interacting generalized AAH model (see Sect.~\ref{sec:AAH}) and is consistent with field theory. 
Within truncated functional RG this can be understood from the effective single-particle SSH model generated during the RG flow. From a perturbative perspective it, in contrast, can be understood as follows. Lowest order perturbation theory for the self-energy, i.e.~the (non-self-consistent) Hartree-Fock approximation, leads to a gap $\sim \delta U$. In the RG process, which includes more than the Hartree-Fock diagrams, this gap is elevated to a power-law with a $U$-dependent exponent.   

We numerically verified that for vanishing average interaction $U=0$ but $\delta U \neq 0$ a gap opens but does not show power-law scaling (in the Hartree-Fock gap) consistent with Eq.~(\ref{eq:gap_dU_scaling_final}).  In short, for $U=0$ the $\delta U$ term alone generates an effective SSH model which shows noninteracting physics.

\begin{figure}[t]
   \centering
   \includegraphics[width=0.5\textwidth]{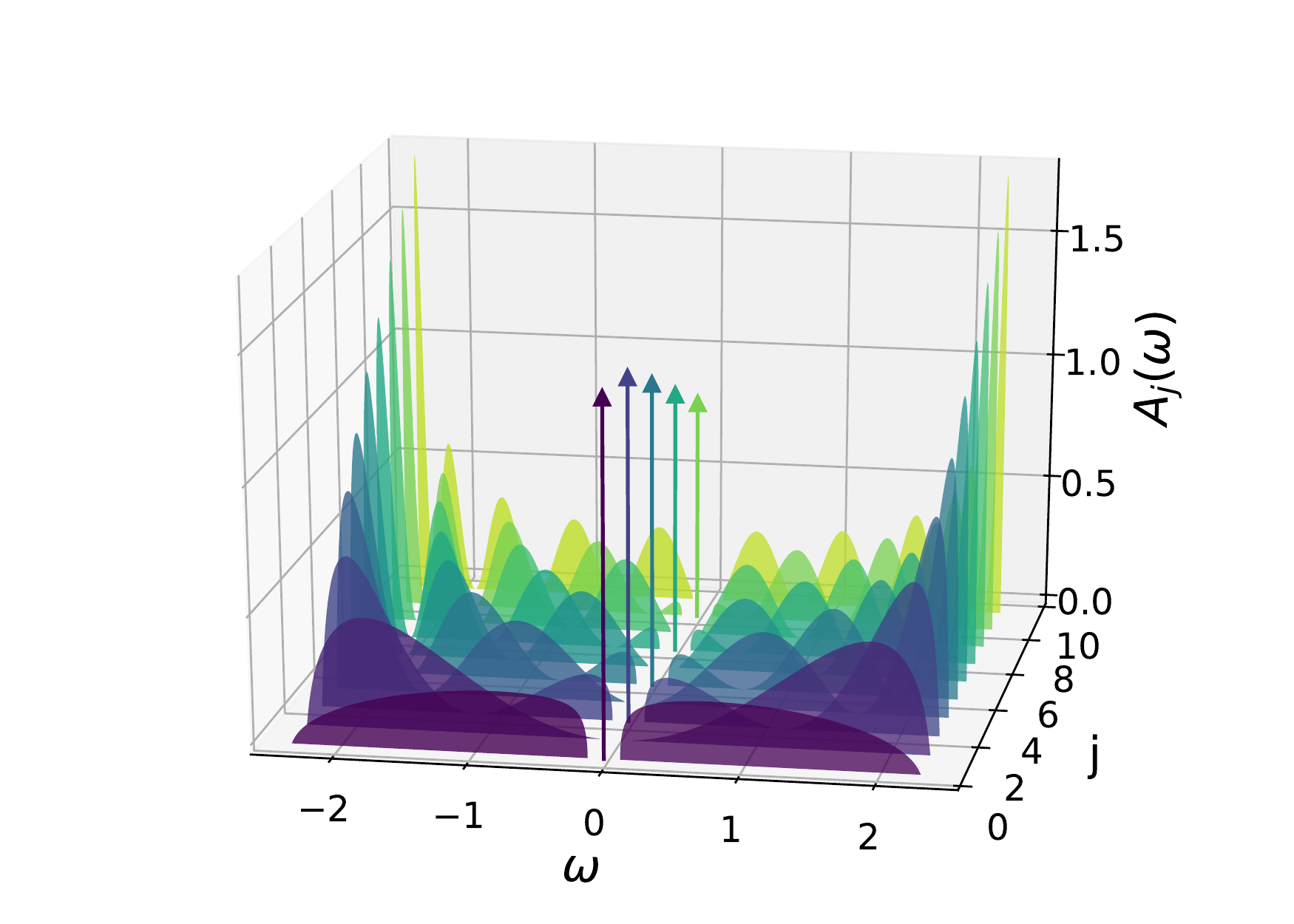}
   \\[0mm]
   \caption{Functional RG data for the local single-particle spectral function $A_j(\omega)$ of the modulated interaction model with $Z=2$, $\delta U=0.19$, $U=0.5$, and $\varphi_U=\pi/4$. Modulated interaction induced in gap zero energy $\delta$-peak are shown as the vertical arrows and the height is scaled up.}
   \label{fig:lsf_u1u2}
\end{figure}

After gaining confidence that also the modulated interaction model can reliably be analyzed using truncated functional RG we proceed and study the local spectral function employing this approach (see our comment of Sect.~\ref{sec:Observable_BSF} on the use of DMRG for this observable). 

In Fig.~\ref{fig:lsf_u1u2} we show functional RG results for the local spectral function $A_j(\omega)$ of the modulated interaction model with $Z=2$ and open boundary conditions as a function of $\omega$ and the lattice site index $j$. The average interaction is $U=0.5$ with a modulation of $\delta U=0.19$. The phase $\varphi_U=\pi/4$ and the system size is given as $N=6000$.
Obviously, an effective edge state forms which shows as an in-gap $\delta$-peak. As the figure indicates the effective edge state has a nontrivial spatial structure. When going from the boundary towards the bulk the spectral weight first increases before its starts to decrease (asymptotically is decreases exponentially; not shown), the latter as it is supposed to be the case for an edge state. 
Due to symmetry the spectral weight of the edge state is only nonvanishing on odd sites (or, equivalently, for the unit cell index $i=1$). 

We now have to ask if this edge state appears because the effective SSH model at the end of the RG flow has bulk single-particle parameters which imply a conventional edge state or if the edge state is of unconventional type. For the effective bulk SSH model an (topological) edge state appears if $t_2^{\rm ren} < t_1^{\rm ren}$ \cite{Su79,Heeger01,RM_frg_prb_20}. For the parameters of Fig.~\ref{fig:lsf_u1u2} this indeed holds and the edge state is of conventional type. It is the well known zero energy edge state of the effective SSH model in its topological phase. The van-Hove-singularity discussed in connection with Fig.~\ref{fig:lsf_aah_z4} as a necessary requirement for the appearance of unconventional edge states, is associated to a modulated onsite energy.
For $Z=2$ and half-filling no (modulated) onsite energy is generated during the RG flow and thus only conventional edge states can be realized. Note that in accordance with property 3.~of the phenomenology described in the Introduction one can still say that the (conventional) edge state is induced by the interaction as the noninteracting model does not have any edge states.     

To summarize, this shows that the $Z=2$ modulated interaction model shows effects similar to the ones summarized in the Introduction. However, due to particle-hole symmetry the density takes its bulk value regardless of the boundary condition.  
To overcome this limitation we next study the $Z=4$ case. 

\subsection{The $Z=4$ model}
\label{subsec:mod_u_Z=4}

From what we just observed for $Z=2$ and our results for the interacting generalized AAH model we expect that the $Z=4$ modulated interaction model will show three gaps corresponding to one-quarter, one-half, and three-quarter filling. Again the Fermi energy is selected such that it lies in the middle of one of the developing gaps. Functional RG but also the Hartree-Fock approximation confirms this.

For $Z=4$ it is less straight forward to determine an analytic expression for the gap at Hartree-Fock level [see Eq.~(\ref{eq:HF_effgap_u1u2}) for $Z=2$]. However, one can easily obtain the Hartree-Fock gap numerically. This shows that $\Delta^{\rm HF}_\nu \sim \delta U$ for one- and three-quarter filling ($\nu=1,3$), that is for the first and the third gap. At half-filling the numerics instead indicates $\Delta^{\rm HF}_{2} \sim \delta U^2$, which is an inconsistent result, as the Hartree-Fock approximation does only include some, but not all, second order contributions for the gap. As discussed in Appendix \ref{App_c} the $\delta U^2$ dependence of the Hartree-Fock gap can be confirmed analytically. We conclude that the special symmetry at half-filling leads to a nongeneric behavior of the corresponding gap. As also in lowest-order truncated functional RG only parts of the second order diagrams for the gap are included this special case can also not be treated in a meaningful way by this method. Half-filling is furthermore difficult to be treated by DMRG. For small $\delta U$, the interesting case when it comes to the power-law scaling, the gap is exceedingly small as indicated by $\Delta^{\rm HF}_{2} \sim \delta U^2$. To access the corresponding low energy regime by DMRG would require huge system sizes  and  very large bond dimensions both beyond reasonable computing resources. We thus refrain from investigating the nongeneric half-filled case any further. It might be a topic for a future study.   

\begin{figure}[t]
   \centering
   \includegraphics[width=0.5\textwidth]{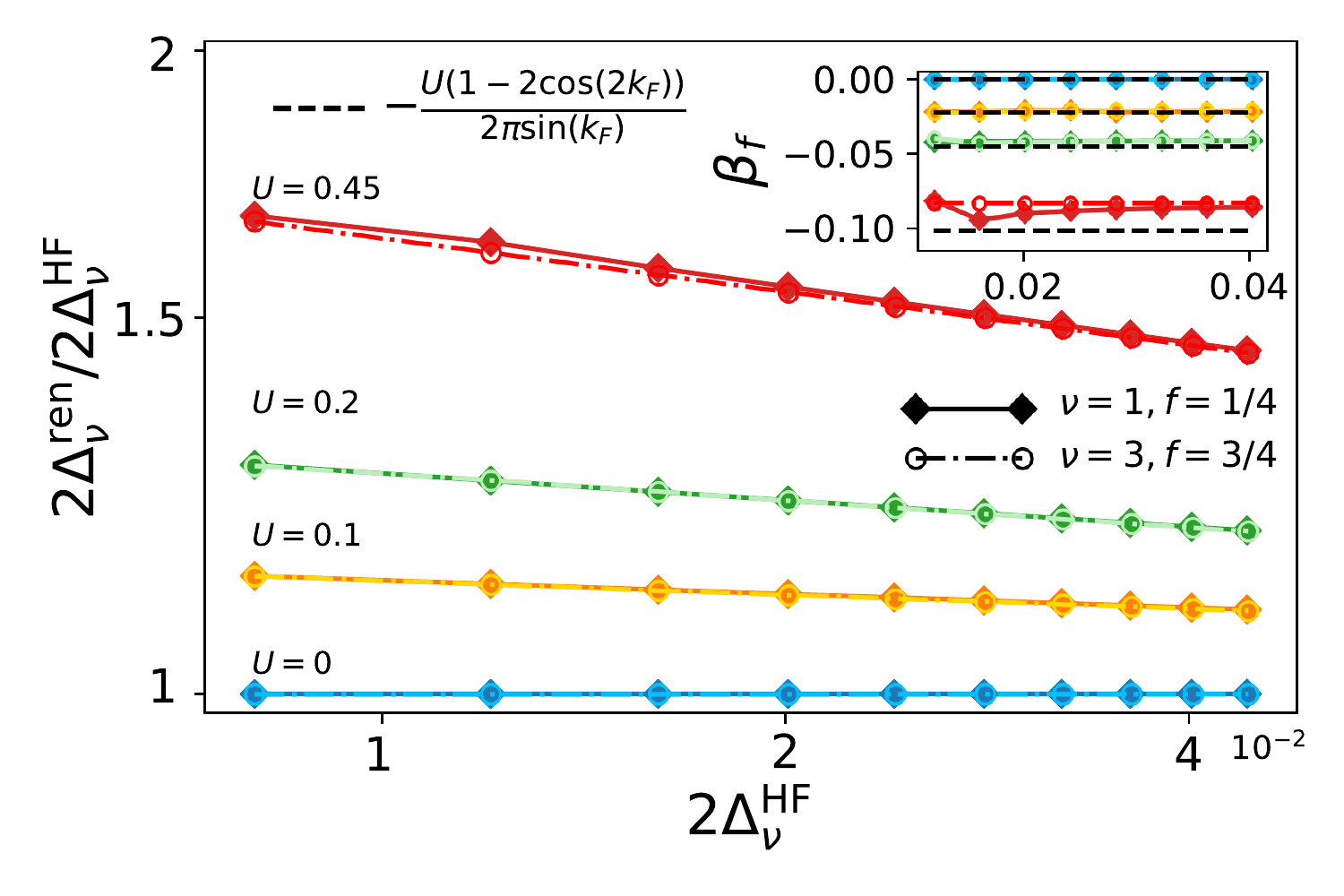}
   \\[0mm]
   \caption{Main panel: The ratio of the functional RG effective gap $2\Delta_{\nu}$ and the effective gap in to first order perturbation theory $2\Delta^{\mathrm{HF}}_{\nu}$ as a function of $2\Delta^{\mathrm{HF}}_{\nu}$. A comparison of quarter
   filling (filled symbols) and three-quarter (open symbols) for different $U$ is shown. The parameters are $U=0.0, 0.1, 0.2, 0.45$, $\varphi_U=-0.45-\pi/4$, the system size is $N=4000$ and $\delta U=0.01, 0.015,\ldots,0.055$. Note the log-log scale. Inset: The log-derivative of the data in the main plot. The dashed line indicates the leading order exponent from field theory.
   }
   \label{fig:eff_gap_mod_u_z4}
\end{figure}

For one- and three-quarter filling we first investigate if Eq.~(\ref{eq:gap_dU_scaling_final}) for the gap also holds at $Z=4$. Based on the finding   that $\Delta^{\rm HF}_\nu \sim \delta U$ (for $\nu=1,3$) and the mechanism leading to the power law as described in the last subsection this appears to be reasonable.
In the main panel of Fig.~\ref{fig:eff_gap_mod_u_z4}, we show functional RG data for $2\Delta_{\nu}/2\Delta^{\text{HF}}_{\nu}$ as a function of $2\Delta^{\text{HF}}_{\nu}$ for different $U$. Data for quarter filling ($\nu=1$) is labeled by filled symbol and for three-quarter filling ($\nu=3$) by open symbols. 
The values of the parameters are given in the caption.
For fixed $U$ the two data sets agree and show linear behavior on a log-log scale. This indicates again the power-law scaling of the effective gap with respect to $2\Delta^{\text{HF}}_{\nu}$ and the same exponent for $\nu=1$ and $\nu=3$. In the inset of Fig.~\ref{fig:eff_gap_mod_u_z4}, we show the centered logarithmic differences of the data in the main panel. For small $2\Delta^{\text{HF}}_{\nu}$ (small $\delta U$), both results approach the leading order exponent $\beta_{f}=-U[1-2\cos (2 k_{\rm F})]/[2\pi\sin(k_{\rm F})]$ from field theory which is indicated as the dashed horizontal lines. As we emphasized in Sec.~\ref{sec:AAH} and our paper on the interacting RM model \cite{RM_frg_prb_20}, we do not control the orders of the exponent beyond the leading one.

Based on these numerical results, we conclude that concerning the gap induced by the modulated interaction the $Z=4$ model at one- and three-quarter filling shows the same behavior as the $Z=4$ interacting generalized AAH model  [see Eq.~(\ref{eq:gap_generic_def})]. In fact, the effective single-particle model at the end of the RG flow of the $Z=4$ modulated interaction model is a AAH model and it appears to be reasonable that both models share the same low energy effective theory. 

 As for $Z=2$ we verified numerically that even for $U=0$ but $\delta U \neq 0$ gaps open but that they do not scale as a power law in the Hartree-Fock gap. The physics of the effective AAH model is that of a noninteracting one.

\begin{figure}[t]
   \centering
   \includegraphics[width=0.5\textwidth]{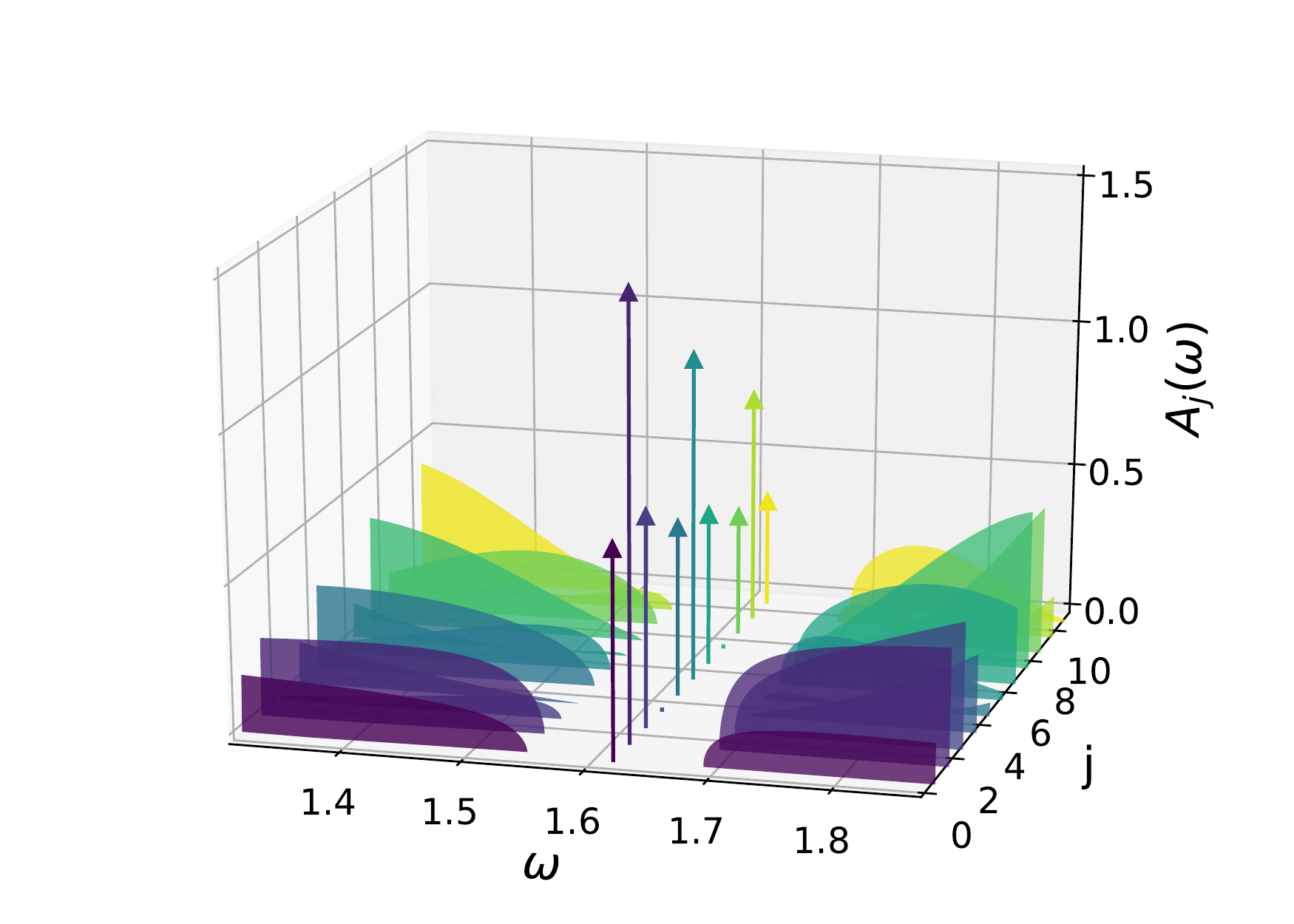}
   \\[0mm]
   \caption{The local single particle spectrum function $A_j(\omega)$ of the three-quarter filling $Z=4$ modulated interaction model as a function of $\omega$ for different lattice sites $j$. The interaction induced effective edge states are indicated by vertical arrows. The parameters are $U=0.25, \delta U=0.15, N=6000, \varphi_U=-\pi/4$.}
   \label{fig:lsf_mod_u_z4}
\end{figure}

Next, we discuss the local spectral function computed by functional RG considering open boundary conditions. In Fig.~\ref{fig:lsf_mod_u_z4} we show $A_j(\omega)$ at three-quarter filling as a function of $\omega$ and the lattice site index $j$. The parameters are given in the caption. The low-energy regime around the Fermi energy which is located in the gap centered around $\omega \approx 1.6$ is shown. The interaction obviously induces an edge state with spectral weight which has a rich spatial structure inherited from the $Z=4$ spatial modulation of the interaction.

As for $Z=2$ the question arises whether the edge state is a conventional or an unconventional one. We verified that for the parameters of Fig.~\ref{fig:lsf_mod_u_z4} the former holds. In contrast to the $Z=2$ case with half-filling the state is, however, not a zero energy edge state. Compared to the interacting generalized AAH model, in which in addition to the interaction $U$, the modulation amplitude and phase of the hopping as well as the onsite energy can be chosen independently the parameter space of the modulated interaction model is significantly smaller. For this only $U$, $\delta U$ and $\varphi_U$ can be chosen. Despite an extensive search of the parameter space we did not succeed to identify a regime in which an unconventional edge state is realized. We, however, emphasize that this does not exclude that such a regime exists. In any case, the (conventional) edge state is induced by the two-particle interaction in accordance with the property 3.~of the Introduction.

\begin{figure}[t]
  \begin{center}
  \includegraphics[width=0.5\textwidth,clip]{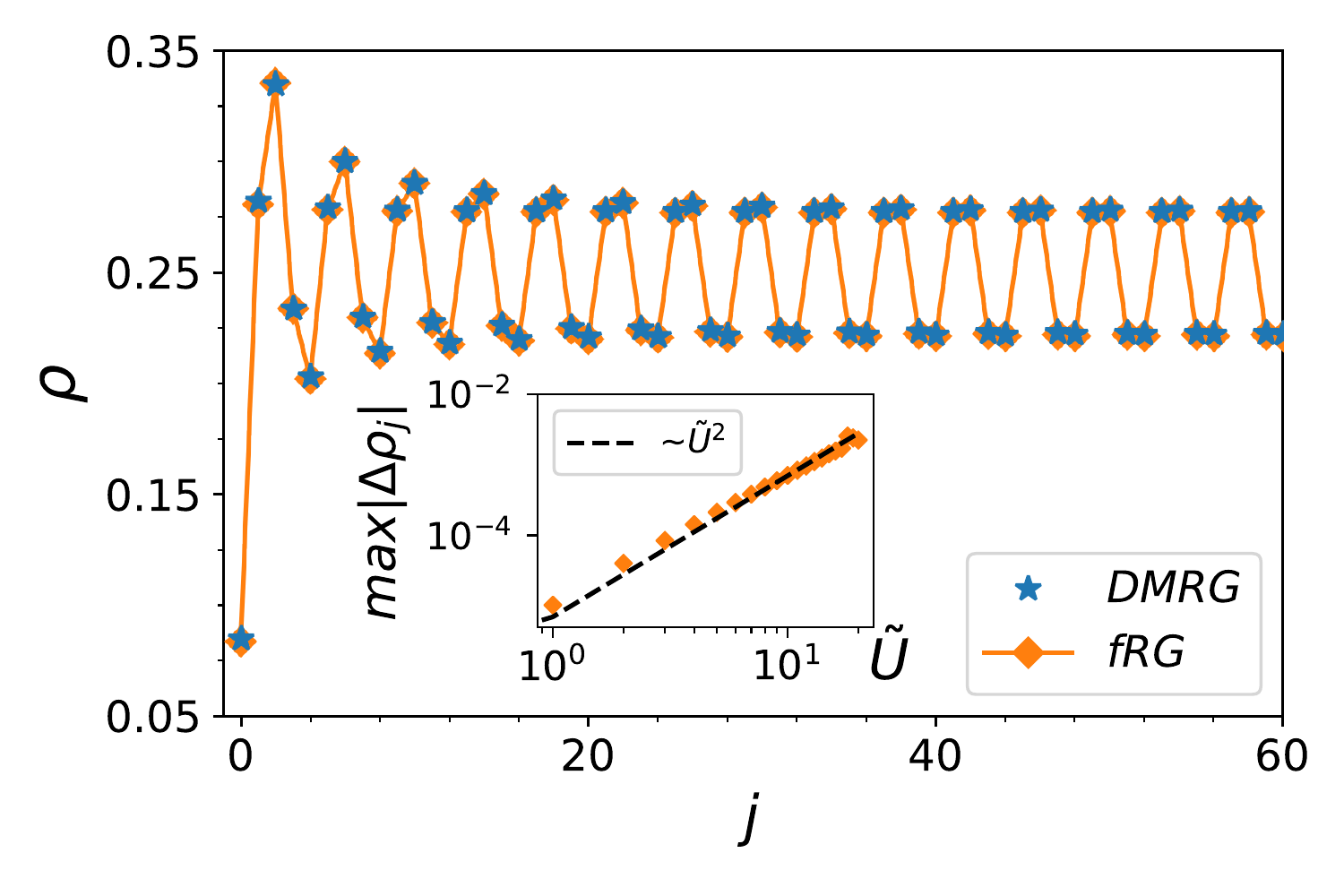}
  \caption{Main plot: Total density $\rho$ as a function of the site index $j$ for quarter filling modulated interaction model with $Z=4, U=-5\delta U=0.375$, and $N=1000$. The resulting functional RG local density which is labeled by ``fRG'' are compared with the DMRG result. Inset:  The largest absolute value of the difference between FRG and DMRG data taken over all of the lattices sites as a function of $\tilde{U}$, where $\tilde{U}=1,2,...,11$ and  $(U,\delta U)=\tilde{U}(0.025,-0.005)$. A log-log scale is used. The dashed line indicated the power law $\tilde{U}^2$.} 
  \label{fig:local_density_FRGvsDMRG_mod_u_z4}
  \end{center}
\end{figure}

We now study the density modulation induced by the modulated interaction for $Z=4$. 
The main panel of Fig.~\ref{fig:local_density_FRGvsDMRG_mod_u_z4} shows a comparison of the local density as a function of lattice site index $j$ computed by functional RG and by DMRG at quarter filling. The parameters can be found in the caption. 
As expected the modulation of the interaction leads to a modulated density around the average value $1/4$. The bulk density, which on the $y$-axis scale of the plot is approximately reached already for $j \approx 50$, reflects the $Z=4$ periodicity. Close to the boundary this periodicity is disturbed by the interplay of the boundary and the (average) interaction. This modulation beyond the unit cell structure decays quickly towards the bulk.

Based on our results for the density of the $Z=4$ interacting generalized AAH model we expect that the decay towards the bulk value is exponential. We verified this (not shown). The decay rate can again be computed by plugging the renormalized bulk parameters at the end of the RG flow into the analytical expression for the noninteracting AAH model. This way the exponential part can be divided out very precisely. 
Based on our insights on the different roles of $\delta U$ and $U$ we anticipate that the pre-exponential function for $U=0$ decays as $1/\sqrt{n}$ (independent of the intra unit cell index $i$) while a more complex, $i$-dependent behavior is found for $U>0$. We also verified this (not shown). The corresponding data look similar to the ones shown for the interacting AAH model in Fig.~\ref{fig:rho_AAH_halffilling}.  

On the scale of Fig.~\ref{fig:local_density_FRGvsDMRG_mod_u_z4} the approximate functional RG data cannot be distinguished from the DMRG ones. In the inset, we show the maximal absolute difference over all lattices sites as a function of $\tilde{U}$, where we define $\tilde{U}$ as $(U,\delta U)=\tilde{U}(0.025,-0.005)$ in order to keep the ratio between $U$ and $\delta U$ constant during the scaling of the interaction. The difference scales as a power-law $\sim \tilde{U}^2$ (dashed line) in full accordance with the approximate nature of the truncated functional RG which does not capture all diagrammatic contributions to second and higher order. 

We found similar results for three-quarter filling corresponding to the gap with index $\nu=3$.

\begin{figure}[t]
   \centering
   \includegraphics[width=0.5\textwidth]{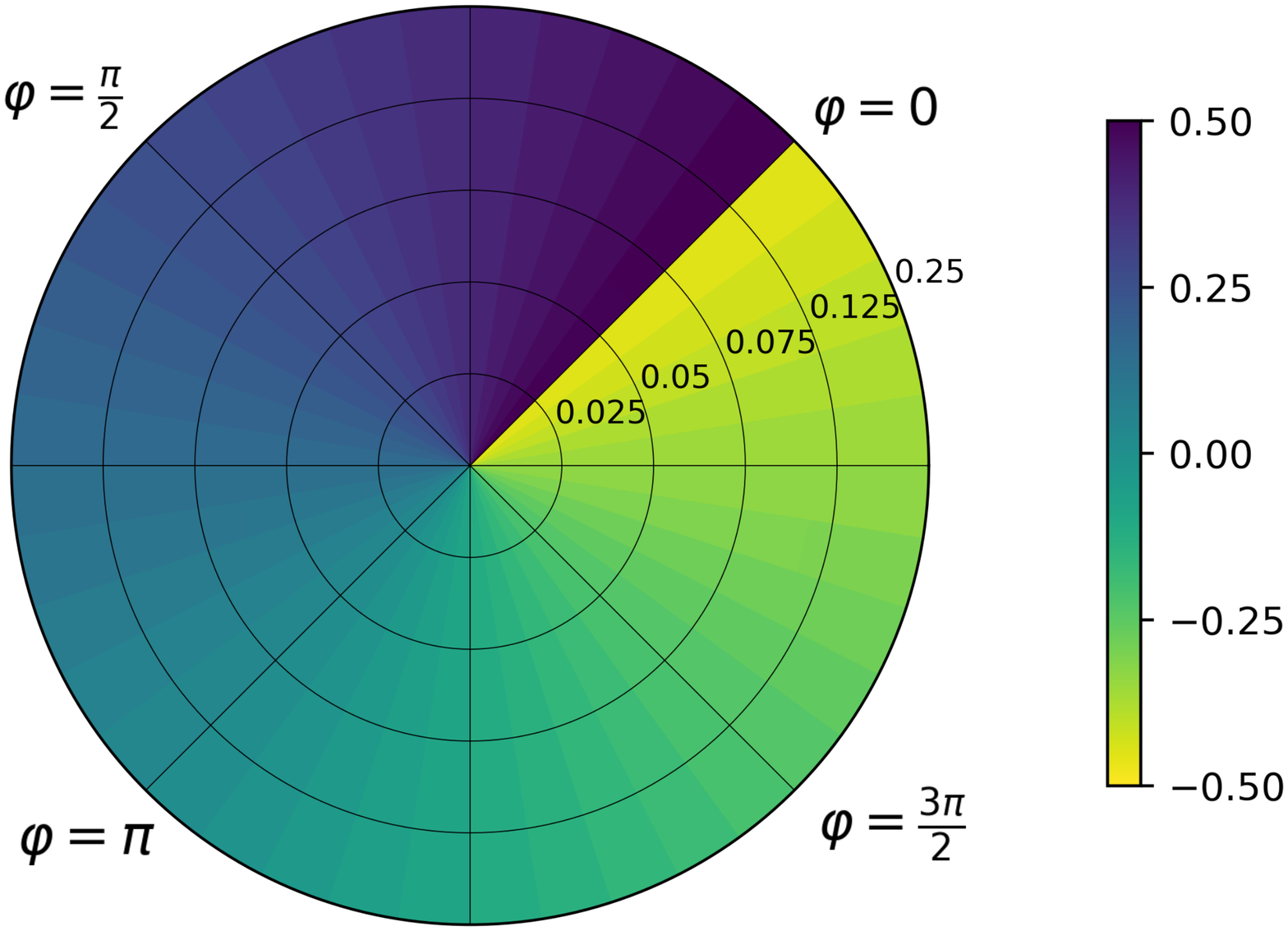}
   \\[0mm]
   \caption{The boundary charge $Q_{\rm B}$ of the three-quarter filled modulated interaction model with $Z=4$ as a function of $\varphi=\varphi_U$ (in the azimuthal direction) and $\delta U$ (in the radial direction), see Eq.~(\ref{eq:QB_mod_u_Z=4}). The system size is given as $N=4000$ and the interaction is $U=0.08$. }.
   \label{fig:qb_mod_u_z4}
\end{figure}

Finally, we investigate the boundary charge of the $Z=4$ modulated interaction model at three-quarter filling. In Fig.~\ref{fig:qb_mod_u_z4} we show functional RG data for the boundary charge as a two-dimensional color-coded plot in polar coordinates. The azimuthal direction displays the angle $\varphi=\varphi_U$ and the radial direction the amplitude of the modulation $\delta U$, which is restricted to small values corresponding to small gaps. The other parameters are given in the caption. We find that $Q_{\rm B}$ depends linearly on $\varphi$ and has a rather weak dependence on $\delta U$. Numerically, we find that the boundary charge is approximately given by 
\begin{align}
    Q_{\rm B} \approx -\frac{\varphi}{2\pi} + \frac{1}{2}.
    \label{eq:QB_mod_u_Z=4}
\end{align}
This result is consistent with Eqs.~(\ref{eq:QB_low_energy}) and (\ref{eq:gap_phase_vs_modulation_phase}) as derived in Refs.~\cite{rational_boundary_charge_in_1d_prr_20,weber_etal_prl_21}. The constant part $1/2$ can be obtained from the special symmetry point $\varphi=\varphi_U=\pi$, where the model has local inversion symmetry such that $Q_{\rm B}=0$.
Our result indicates that $Q_{\rm B}$ shows the same behavior as for the interacting AAH model for small gaps; see Sec.~\ref{sec:AAH}. This is consistent with the observation from the other observables that the modulated interaction and the AAH model share the same effective low-energy theory.
We, moreover, note (without showing data) that one can use the noninteracting AAH model with the bare bulk parameters replaced by the renormalized effective bulk parameters to generate the same result. This supports the idea that the characteristics of the boundary charge  can  fully be determined by the bulk properties with effective single-particle parameters. This is summarized by the property 4.~of the introduction. 
We found similar results for one-quarter filling.

\section{Summary}
\label{sec:summary}

This work generalizes the results of Ref.~\cite{RM_frg_prb_20} on interaction effects in the $Z=2$ RM model to  microscopic lattice models with more complex unit cell structure. A posteriori the success of this generalization also lends credence to similar arguments obtained from a purely field theoretical approach.  


We first investigated a direct extension of the $Z=2$ RM model, namely, an interacting generalized 1d AAH model with a unit cell of size $Z=4$. We employed functional RG, which proved to be a valuable tool in the prior study of the RM model. We were able to confirm the phenomenology already identified for the interacting RM model also in the case $Z=4$. The gap is renormalized by the two-particle interaction. For small bare gaps the renormalized one scales as a power law with the two-particle interaction entering the exponent. The local spectral function close to an open boundary can show (depending on the single-particle parameters) an interaction induced in-gap $\delta$-peak associated to an effective edge state. Its appearance cannot be explained based on the bare or renormalized bulk single-particle parameters. In contrast, the characteristics of the boundary charge accumulated close to an open boundary are robust towards the two-particle interaction. This consolidates our idea that the boundary charge might provide a more robust relation between bulk and boundary properties than the number of edge states (conventional bulk-boundary correspondence). 

In a second step we investigated a lattice model which in the noninteracting limit is translationally invariant by a single-lattice site (for PBCs). In this the non-trivial unit cell structure is induced by a periodically modulated two-particle interaction. Up to singular situations with a special symmetry this model also followed the above discussed phenomenology, with the difference being that the interaction induced effective edge states can be predicted based on the renormalized single-particle parameters. For the parameter space we explored, we did not find any interaction induced edge states which cannot be explained this way. 

\section*{Acknowledgments}
This work was supported by the Deutsche Forschungsgemeinschaft  (DFG, German Research Foundation) via RTG 1995 and under Germany's Excellence Strategy - Cluster of Excellence Matter and Light for Quantum Computing (ML4Q) EXC 2004/1 - 390534769. DMK acknowledges support from the Max Planck-New York City Center for Non-Equilibrium Quantum Phenomena. Simulations were performed with computing resources granted by RWTH Aachen University.

\appendix

\section{The flow equations }
\label{App_a}

In this Appendix we present the truncated functional RG equations for the self-energy and the density for arbitrary $Z$. They hold for both the interacting generalized AAH model as well as the modulated interaction models. We start out with the real space equations which can be employed for open as well as periodic boundary conditions.

\subsection{Real space}
\label{App_a_real}

We focus on the lowest-order truncated functional RG scheme featuring a static flowing self-energy  $\Sigma^{\Lambda}$
and consider a sharp frequency cutoff in Matsubara space \cite{RM_frg_prb_20,Andergassen04}.
The two-particle vertex function in real space is given as 
\begin{align}
  &\Gamma_{j'_1,j'_2;j^{}_1,j^{}_2} = U_{j_1,j_2} \left(\delta_{j^{}_1,j'_1}\delta_{j^{}_2,j'_2}- \delta_{j^{}_1,j'_2}\delta_{j^{}_2,j'_1}\right), \\
  &U_{j^{}_1,j^{}_2}    = U_{j^{}_1}\delta_{j^{}_1,{j^{}_2-1}}+U_{j^{}_1-1}\delta_{j^{}_1,j^{}_2+1},
\end{align}
and the lowest-order flow equation of the self-energy reads
\begin{align}
  \partial_{\Lambda} \Sigma^{\Lambda}_{j'_1,j^{}_1} &=-\frac{1}{2\pi}\sum_{\omega=\pm\Lambda}e^{i\omega 0^{+}} \mathcal{G}^{\Lambda}_{j^{}_2,j'_2} (i \omega) \Gamma_{j'_1,j'_2;j^{}_1,j^{}_2} \\
  	&=-\frac{1}{\pi}\text{Re}\left[  \mathcal{G}^{\Lambda}_{j^{}_2,j'_2} (i \Lambda)  U_{j_1,j_2} \delta_{j^{}_1,j'_1}\delta_{j^{}_2,j'_2} \right] \label{eq:AA}\\
  	& \hspace{1.em} +\frac{1}{\pi}\text{Re}\left[  \mathcal{G}^{\Lambda}_{j^{}_2,j'_2} (i \Lambda)  U_{j_1,j_2}  \delta_{j^{}_1,j'_2}\delta_{j^{}_2,j'_1}\right]\label{eq:BB}
  \end{align}
with a cutoff dependent propagator
\begin{align}
  {\mathcal G}^{\Lambda}({\rm i}\omega) =
  \left\{ \left[ {\mathcal G}_{0}({\rm i}\omega) \right]^{-1} -
  \Sigma^{\Lambda} \right\}^{-1}.
  \label{eq:SSP}
\end{align}	
Equations (\ref{eq:AA}) and (\ref{eq:BB}) are the flow equations for the diagonal part of the self-energy and the off-diagonal one respectively. They can be brought in the more explicit forms
\begin{align}
&\partial_{\Lambda} \Sigma^{\Lambda}_{j,j} = -\frac{1}{\pi}\text{Re}\left[  \mathcal{G}^{\Lambda}_{j+1,j+1} (i \Lambda)  U_{j} + \mathcal{G}^{\Lambda}_{j-1,j-1} (i \Lambda)U_{j-1} \right]. \\
&\partial_{\Lambda} \Sigma^{\Lambda}_{j,j\pm1} =\frac{1}{\pi}\text{Re}\left[  \mathcal{G}^{\Lambda}_{j,j+1} (i \Lambda)  U_{j/j-1} \right] 
\end{align}	

To consistently compute the local density $\rho(j)$ we set up according
flow equations for this observable 
\begin{align}
  &\frac{\partial}{\partial \Lambda} \rho^{\Lambda}(j) = -\frac{1}{2\pi}
    \sum_{\omega=\pm \Lambda} \mbox{tr} \left[ e^{\mbox{i}\omega 0^{+} }
    {{\mathcal G}}^{\Lambda}({{\rm i}\omega})R^{\Lambda}_{j}({{\rm i}\omega})
    \right] .
  \label{eq:RG_density_eq}
\end{align}
They involve a density response vertex $R^{\Lambda}_{j}$ which obeys the equations 
\begin{widetext}
\begin{align}
   \frac{\partial}{\partial \Lambda} R^{\Lambda}_{j;l,l} &= -\frac{1}{2\pi} 
   \sum_{\omega=\pm \Lambda}
      \sum_{l'}\hspace{-0.5em}\sum_{r^{'}=0,\pm1}  U_{l}  {{\mathcal G}}^{\Lambda}_{l+1,l'}({{\rm i}\omega}) R^{\Lambda}_{j;l',l'+r'}{{\mathcal G}}_{l'+r',l+1}^{\Lambda}({{\rm i}\omega})  
     	+U_{l-1}  {{\mathcal G}}^{\Lambda}_{l-1,l'}({{\rm i}\omega}) R^{\Lambda}_{j;l',l'+r'}{{\mathcal G}}_{l'+r',l-1}^{\Lambda}({{\rm i}\omega}),
                                                      \\  
  \frac{\partial}{\partial \Lambda} R^{\Lambda}_{j;l,l\pm1}&= -\frac{1}{2\pi} 
   \hspace{-0.5em} \sum_{\omega=\pm \Lambda} 
    \sum_{m m'}U_{l/ l-1}  {{\mathcal G}}^{\Lambda}_{l,m'}({{\rm i}\omega}) 
                                        R^{\Lambda}_{j;m',m}
                                          {{\mathcal G}}_{m,l\pm1}^{\Lambda}({{\rm i}\omega}).
\end{align}
\end{widetext}
Details on the advantage of computing $\rho(j)$ via its own flow equation as compared to an approach employing the Green function (and thus the self-energy) can be found in Ref.~\cite{RM_frg_prb_20}.

Similar to what is done in Ref.~\cite{RM_frg_prb_20}, one can decompose the self-energy into unit cell index independent and dependent parts, and the renormalized parameters at the end of the RG prodedure are given as
\begin{align}
&V^{\rm ren}_{j=Z(n-1)+i}=V_j+\Sigma^{\Lambda=0}_{j,j}=V^{\rm ren}_i+V^{\rm F}_i(n), \\
&t^{\rm ren}_{j=Z(n-1)+i}=t_j+\Sigma^{\Lambda=0}_{j,j+1}=t^{\rm ren}_i+t^{\rm F}_i(n).
\end{align}
where ``ren'' and ``F'' denotes the unit cell index $n$ independent and dependent parts of the effective onsite potential and the hopping parameters. 

The above sets of equations can for large but finite systems easily be solved on a computer.

\subsection{Momentum space}
\label{App_a_momentum}

For underlying periodic boundary conditions on can derive flow equations for the effective single particle parameters in $k$-space. They are set up directly in the thermodynamic limit. For the hopping parameters  one obtains 
\begin{align}  
  &\partial_\Lambda t^{\Lambda}_i=-\frac{U^{}_i}{2\pi} \sum_{\omega=\pm\Lambda}\int^{\pi}_{-\pi}\frac{dk}{2\pi} \mathcal{G}^{\Lambda}_{i,i+1}(k,i\omega)  \text{ for $i \neq Z$}, \nonumber \\ 
  &\partial_\Lambda t^{\Lambda}_{i=Z}=-\frac{U^{}_Z}{2\pi} \sum_{\omega=\pm\Lambda}\int^{\pi}_{-\pi}\frac{dk}{2\pi} e^{i k}\mathcal{G}^{\Lambda}_{Z,1}(k,i\omega) \text{ for  $i=Z$}. 
  \label{eq:tflow_imp}
 \end{align}
 For the onsite potential they read
\begin{widetext}
\begin{align}  
  &\partial_\Lambda V^{\Lambda}_{i=1}=-\frac{1}{2\pi} \sum_{\omega=\pm\Lambda}\int^{\pi}_{-\pi}\frac{dk}{2\pi} \left\{ U^{}_1\mathcal{G}^{\Lambda}_{2,2}(k,i\omega)+U^{}_Z\mathcal{G}^{\Lambda}_{Z,Z}(k,i\omega) \right\}  \text{ for  $i=1$}, \nonumber \\
  &\partial_\Lambda V^{\Lambda}_{i=Z}=-\frac{1}{2\pi} \sum_{\omega=\pm\Lambda}\int^{\pi}_{-\pi}\frac{dk}{2\pi} \left\{ U^{}_{Z}\mathcal{G}^{\Lambda}_{1,1}(k,i\omega)+U^{}_{Z-1}\mathcal{G}^{\Lambda}_{Z-1,Z-1}(k,i\omega) \right\} \text{ for  $i=Z$}, \nonumber \\  
  &\partial_\Lambda V^{\Lambda}_i=-\frac{1}{2\pi} \sum_{\omega=\pm\Lambda}\int^{\pi}_{-\pi}\frac{dk}{2\pi} \left\{ U_i\mathcal{G}^{\Lambda}_{i+1,i+1}(k,i\omega)+U^{}_{i-1}\mathcal{G}^{\Lambda}_{i-1,i-1}(k,i\omega) \right\} \text{ else}.  \nonumber \\ 
 \end{align}
 and the single scale propagator in the right hand side of the RG equations is given as
\begin{align}
\{\mathcal{G}^{\Lambda}(k,i\omega)\}^{-1} =
\left( \begin{array}{rrrr}
i\Lambda+\mu-V^{\Lambda}_1 & t^{\Lambda}_1\phantom{--} &    & t^{\Lambda}_Ze^{-ik}\phantom{-} \\
t^{\Lambda}_1\phantom{--} & i\Lambda+\mu-V^{\Lambda}_2 & \ddots & \phantom{--}  \\
\phantom{--} & \ddots\phantom{--} & \ddots & t^{\Lambda}_{Z-1}\phantom{--}  \\
t^{\Lambda}_Ze^{ik}\phantom{-} &  & t^{\Lambda}_{Z-1} & i\Lambda+\mu-V^{\Lambda}_Z \\
\end{array}\right) .
\end{align}
 \end{widetext}

The above set can be easily be solved numerically. However, it can also be used as the starting point for an analytical analysis; see Appendix \ref{App_b} for the $Z=2$ modulated interaction model.   
 
\section{Analytical insights for the modulated interaction model with $Z=2$ }
\label{App_b}

Here we consider the $Z=2$ modulated interaction model with vanishing Fermi energy, i.e.~at half filling.
After plugging in the propagator and performing the $k$-integration in the right hand side of the RG equation (\ref{eq:tflow_imp}), one obtains 
\begin{align}
	&\partial_{\Lambda} 2\Delta^{\Lambda}=  2\Delta^{\Lambda}\frac{U}{\pi} \frac{1}{b_{}^{\Lambda}} \left\{ 1- \frac{a^{\Lambda}+b^{\Lambda}}{\sqrt{(a^{\Lambda})^2-(b^{\Lambda})^2}}\right\} \nonumber  \\
	&\hspace{4.5em}-2W^{\Lambda}\frac{\delta U}{\pi} \frac{1}{b_{}^{\Lambda}} \left\{ 1- \frac{a^{\Lambda}-b^{\Lambda}}{\sqrt{(a^{\Lambda})^2-(b^{\Lambda})^2}}\right\},  \label{eq:RG_gap_eq} \\
	&\partial_{\Lambda} 2W^{\Lambda}=  -2W^{\Lambda}\frac{U}{\pi} \frac{1}{b_{}^{\Lambda}} \left\{ 1- \frac{a^{\Lambda}-b^{\Lambda}}{\sqrt{(a^{\Lambda})^2-(b^{\Lambda})^2}}\right\} \nonumber  \\
	&\hspace{4.5em}+2\Delta^{\Lambda}\frac{\delta U}{\pi} \frac{1}{b_{}^{\Lambda}} \left\{ 1- \frac{a^{\Lambda}+b^{\Lambda}}{\sqrt{(a^{\Lambda})^2-(b^{\Lambda})^2}}\right\} \label{eq:RG_bandwidth_eq}
\end{align}
for the difference and the sum of the two hoppings $t_1^\Lambda$ and $t_2^\Lambda$, respectively. Here
$a_{}^{\Lambda}=\Lambda^2+(t^{\Lambda}_{1})^2+(t^{\Lambda}_{2})^2$, and $b_{}^{\Lambda}=2t^{\Lambda}_{1}t^{\Lambda}_{2}$.

From these flow equations one can obtain the effective gap and bandwidth in first order perturbation theory, i.e.~on Hartree-Fock level, by turning off the feedback of self-energy in the right hand side of Eq.~(\ref{eq:RG_gap_eq}) and~(\ref{eq:RG_bandwidth_eq}) \cite{RM_frg_prb_20}. The $\Lambda$-integral can then be performed leading to 
\begin{align}
	&2\Delta^{\Lambda=0}=2\Delta^{\text{HF}}= \frac{4\delta U}{\pi},  \label{eq:HF_gap} \\
	&2W^{\Lambda=0}=2W^{\text{HF}}=  4t\left(1+\frac{U}{\pi t}\right),\label{eq:HF_bandwidth}
\end{align}
for the Hartree-Fock gap $2\Delta^{\text{HF}}$ and bandwidth $2W^{\text{HF}}$.
For $\Delta^{\Lambda}\ll\Lambda,W^{\Lambda}$, that is in the small gap  limit, and keeping terms to linear order in  $U$ and $\delta U$ on the right hand side of  Eq.~(\ref{eq:RG_gap_eq}), one 
finds
\begin{align}
	&\partial_{\Lambda} \Delta^{\Lambda}= \frac{1}{2t^2}\left\{ \Delta^{\Lambda}\frac{U}{\pi}\left(1-\frac{\sqrt{\Lambda^2+4t^2}}{\Lambda}\right) \right. \nonumber \\
&& \left. \hspace{-20em}-2t\frac{\delta U}{\pi}\left(1-\frac{\Lambda}{\sqrt{\Lambda^2+4t^2}}\right)\right\}. \nonumber  
\end{align}

\begin{figure}[t]
   \centering
   \includegraphics[width=0.5\textwidth]{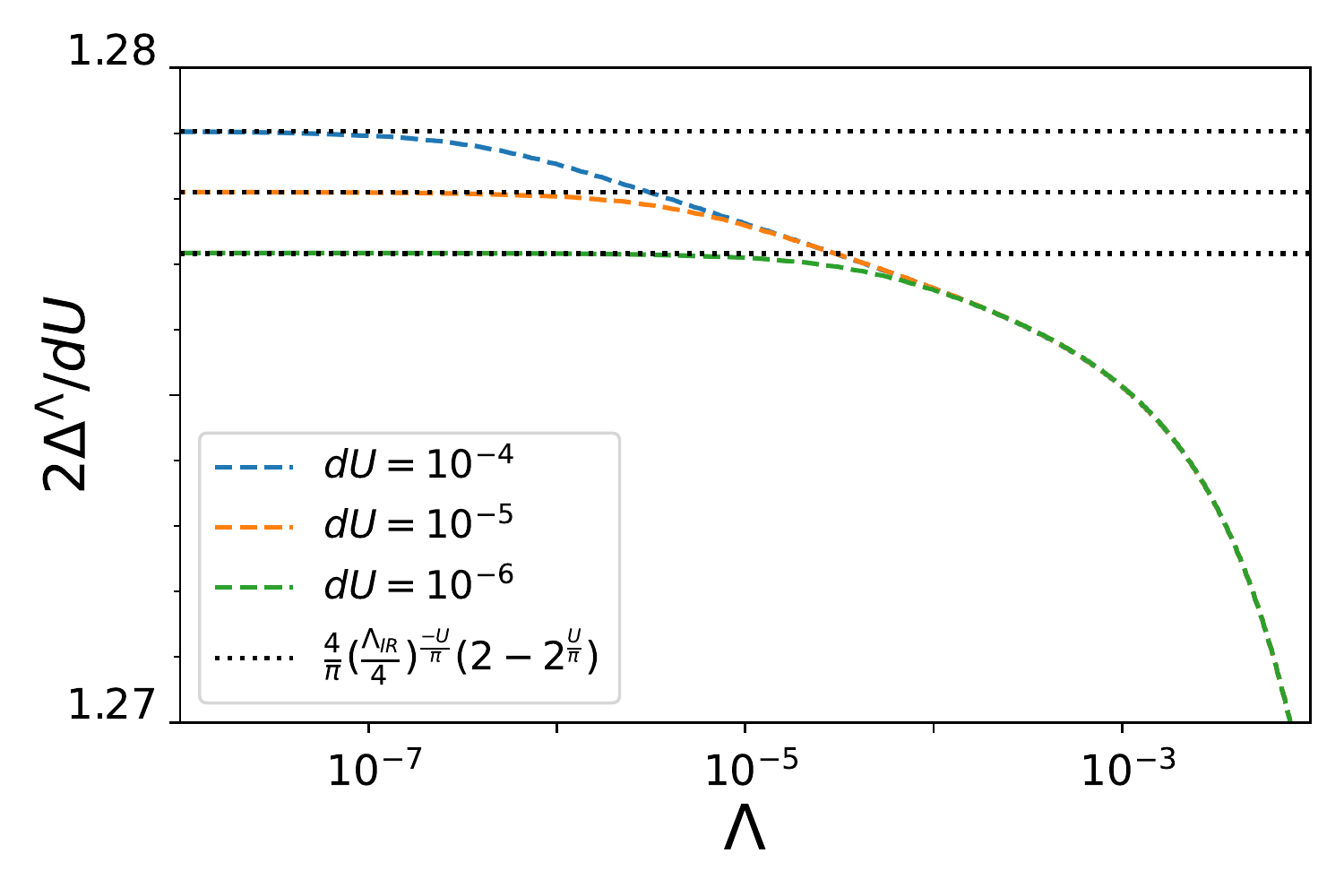}
   \\[0mm]
   \caption{Exemplary numerical data for the $k$-space RG flow (color dashed lines) and the analytical result [Eq.~(\ref{eq:gap_dU_scaling}) with Eq.~(\ref{eq:IR_cutoff}) substituted; black dotted lines] of the ratio of effective gap and $\delta U$ are shown. The model parameters are given as $U=0.001$, $t=1$ and different $\delta U$ are shown in the plot. The infrared cutoff scale is clearly visible.}
   \label{fig:flow_gap_u1u2_smallU}
\end{figure}

One can rescale all of the parameters $\Delta^{\Lambda}\rightarrow y=\Delta^{\Lambda}/2t$, $\Lambda\rightarrow x=\Lambda/2t$, $U\rightarrow a=U/{\pi t}$ and $\delta U\rightarrow c=\delta U/{\pi t}$,
to obtain the dimensionless equation
\begin{align}
	&\frac{dy}{dx}= a y \left(1-\frac{\sqrt{1+x^2}}{x}\right) -c\left(1-\frac{1}{\sqrt{1+x^2}}\right) ,\label{eq:RG_gap_dimensionless} 
\end{align}
with the initial condition $y(x=\infty)=0$. The solution of Eq.~(\ref{eq:RG_gap_dimensionless}) is given as 
\begin{align}
	y(x)= c &\exp\left[a(x-\sqrt{x^2+1})\right] \left(\frac{x}{1+\sqrt{1+x^2}}\right)^{-a}\times \nonumber  \\
	& \hspace{-2em} \int^x_{\infty}d\xi \left\{ \exp \left[a(\xi-\sqrt{\xi^2+1})\right] \left(\frac{\xi}{1+\sqrt{1+\xi^2}}\right)^{a} \right.  \nonumber \\
& \left. \hspace{1.5em}  \left(\frac{\xi-\sqrt{1+\xi^2}}{\sqrt{1+\xi^2}}\right) \right\}
	 \label{eq:solution_dimensionless} 
	\end{align}
In order to obtain the result in the small $a$ and $c$ limit (small $\Delta U$ and $U$ limit), one keeps the linear in $c$ term in front of the expression and the linear in $a$ term in the exponent. Moreover, in this limit one finds $\exp\left[a(x-\sqrt{x^2+1})\right]\cong1$ and $\exp\left[a(\xi-\sqrt{\xi^2+1})\right]\cong1$. Therefore,  
\begin{align}
	y(x)
	\cong & c \left(\frac{x}{1+\sqrt{1+x^2}}\right)^{-a} \frac{2\left(\tan \frac{\xi}{2}+1\right)^{a-1}}{(a-1)} \times \nonumber  \\ & \hspace{1.5em}\left._2F_1\left(1-a,-a;2-a;\frac{1}{1+\tan(\frac{\xi}{2})}\right) \right|^{\xi=\frac{\pi}{2}}_{\xi=\tan^{-1}x} \nonumber  \\
	\cong & c \left(\frac{x}{2}\right)^{-a}\left(2-2^a\right)   \hspace{1.5em}\mbox{for $x=\frac{\Lambda}{2t}\ll1$}
	 \label{eq:solution_smallU} 
	\end{align}
Finally, the effective gap for small two-particle interactions is given as
\begin{align}
	2\Delta^{\text{eff}}= \frac{4 \delta U}{\pi} \left(\frac{\Lambda_{\text{IR}}}{4t}\right)^{-U/\pi t}(2-2^{U/\pi t}),	 \label{eq:gap_dU_scaling}
	\end{align}
where $\Lambda_{\text{IR}}$ is the IR cutoff of the flow. From numerical observation, we find
\begin{align}
	\Lambda_{\text{IR}}=2\Delta^{\text{HF}}=\frac{4 \delta U}{\pi},\label{eq:IR_cutoff} 
	\end{align}
which is the effective gap in first order perturbation theory, see Eq.~(\ref{eq:HF_gap}). For an exemplary parameter set this is illustrated in Fig.~\ref{fig:flow_gap_u1u2_smallU}.

\section{The modulated interaction model with $Z=4$ at half filling}
\label{App_c}

In this Appendix, we investigate the effective gap in lowest order perturbation theory (for the self-energy) for the modulated interaction model with $Z=4$ at half filling. For this we can use the flow equations in momentum space.
Due to the particle hole symmetry, the flow of the on-site potential vanishes. The flow of the effective hopping parameters are given as 
\begin{align}  
  &\partial_\Lambda t^{\Lambda}_1=-\frac{U_1}{2\pi} \sum_{\omega=\pm\Lambda}\int^{\pi}_{-\pi}\frac{dk}{2\pi} \mathcal{G}^{\Lambda}_{1,2}(k,i\omega),  \nonumber \\ 
  &\partial_\Lambda t^{\Lambda}_2=-\frac{U_2}{2\pi} \sum_{\omega=\pm\Lambda}\int^{\pi}_{-\pi}\frac{dk}{2\pi} \mathcal{G}^{\Lambda}_{2,3}(k,i\omega), \nonumber \\  
  &\partial_\Lambda t^{\Lambda}_3=-\frac{U_2}{2\pi} \sum_{\omega=\pm\Lambda}\int^{\pi}_{-\pi}\frac{dk}{2\pi} \mathcal{G}^{\Lambda}_{3,4}(k,i\omega), \nonumber \\ 
  &\partial_\Lambda t^{\Lambda}_4=-\frac{U_4}{2\pi} \sum_{\omega=\pm\Lambda}\int^{\pi}_{-\pi}\frac{dk}{2\pi} e^{-i k}\mathcal{G}^{\Lambda}_{4,1}(k,i\omega),
 \end{align}
 Without the feedback of the self-energy in the single scale propagator, that is on Hartree-Fock level,
 the effective hopping parameters at the end of the flow are given as 
\begin{align}
t^{\text{HF}}_i & = t^{\Lambda=0}_i\nonumber \\ &=t-\frac{U_i}{\pi} \int^{0}_{\infty}d\Lambda \int^{\pi}_{-\pi}\frac{dk}{2\pi} \frac{t \left[\Lambda^2+t^2(1-\cos(k))\right]}{ \Lambda^4+4t^2\Lambda^2+2t^4(1-\cos(k))} \nonumber \\
&=t-\frac{U_i}{\pi}\int^{0}_{\infty}d\Lambda  \frac{1}{2t}\left\{ 1- \frac{\Lambda}{\sqrt{\Lambda^2+(2t)^2}}\right\} \nonumber \\ 
&=t+\frac{U_i}{\pi}
\end{align}
Without loss of generality, we choose $\varphi_U=\pi/2$ and set  
\begin{align}
	&t^{\text{HF}}_1=t^{\text{HF}}_3=t+\frac{U}{\pi} = \bar{t},\\
	&t^{\text{HF}}_{2/4}=t+\frac{U\pm \delta U}{\pi}= \bar{t}\pm\delta\bar{U}.
\end{align}
 With the  effective single-particle Hamiltonian in momentum space $h_k$ and the dispersion relation $E(k)$, one can rewrite the eigenvalue equation $\det[h_k-E(k)]=0$ into 
\begin{align}
2\bar{t}^2 (\bar{t}^2-\delta\bar{U}^2)\cos k= E^4(k)-AE^2(k)+B,
\end{align}
where $A=4\bar{t}^2+2\delta\bar{U}^2$ and $B=\delta\bar{U}^4$. Therefore, the band energy is given as 
\begin{align}
E(k)=\pm\sqrt{\frac{A\pm\sqrt{{A^2-4(B-2\bar{t}^2 (\bar{t}^2-\delta\bar{U}^2)\cos k)}}}{2}}
\end{align}
The gap is located at $k=0$ and 
\begin{align}
2\Delta^{\text{HF}}_{\nu=2}
= & 2\sqrt{2 \bar{t}^2+\delta\bar{ U}^2-2\bar{t}^2\sqrt{1+\frac{\delta\bar{ U}^2}{\bar{t}^2}} }\\
\cong & \frac{\delta U^2}{t\pi^2}.
\end{align}
We conclude, that even with the linear $U$ and $\delta U$ terms in the effective hopping parameters, the linear term in the effective gap is cancelled out. This shows that the effective gap of the modulated interaction model with $Z=4$ at half filling is at least of second order in the interaction. In our lowest order truncated functional RG, not all of the second order terms are taken into account. Therefore, we can not obtain controlled results within lowest-order truncated functional RG.


\begin{thebibliography}{}
\bibitem{com}
The bulk polarization discussed in Refs.~\cite{resta_prl_98,resta_sorella_prl_99} can be related to the boundary charge via the so-called surface charge \cite{kingsmith_vanderbilt_prb_93,pletyukhov_kennes_klinovaja_loss_schoeller_prb_20} and surface fluctuation theorem \cite{weber_etal_prl_21}.
\end{thebibliography}
\end{document}